\documentclass[longauth]{aa}

\def\grvs{\ensuremath{G_{\rm RVS}}\xspace}
\def\obgrvs{\ensuremath{G_\mathrm{RVS}^{\rm{onboard}}}\xspace}
\def\extgrvs{\ensuremath{G_\mathrm{RVS}^{\rm{ext}}}\xspace}

\def\refgrvs{\ensuremath{G_\mathrm{RVS}^{\rm{ref}}}\xspace}
\def\epochgrvs{\ensuremath{G^{\rm{epoch}}_\mathrm{RVS}}\xspace}
\def\grpgrvs{\ensuremath{G^{\rm{G,RP}}_\mathrm{RVS}}\xspace}

\def\gbp{\ensuremath{G_{\rm BP}}\xspace}
\def\grp{\ensuremath{G_{\rm RP}}\xspace}
\def\g{\ensuremath{G}\xspace}

\newcommand\gaia{\textit{Gaia}}

\newcommand\kms{\ensuremath{\text{km~s}^{-1}}}

\newcommand{\es}{$\mathrm{e}^{-}\mathrm{s}^{-1}$}
\def\arcsec{\ensuremath{''}}

\def\simgt{\mathrel{\hbox to 0pt{\lower 3.5pt\hbox{$\mathchar"218$}\hss}
    \raise 1.5pt\hbox{$\mathchar"13E$}}}
\def\simlt{\mathrel{\hbox to 0pt{\lower 3.5pt\hbox{$\mathchar"218$}\hss}
    \raise 1.5pt\hbox{$\mathchar"13C$}}}

\usepackage{graphicx}
\usepackage{txfonts}

\usepackage{amsmath}
\usepackage{hyperref}
\begin{document} 

\newcommand\vtidy{\vspace{-\baselineskip}}
\newcommand\etc{{\it etc.\ }}
\newcommand\eg{e.g.\ }
\newcommand\ie{i.e.\ }

\title{{\gaia} Data Release 3\\ \grvs\ photometry from the RVS spectra}

%\subtitle{}
\author{
%managers and old
P.~Sartoretti                    \inst{\ref{gepi}}  
\and O.      ~Marchal                            \inst{\ref{strasbourg}} 
\and C.      ~Babusiaux                            \inst{\ref{grenoble},\ref{gepi}} 
\and C.      ~Jordi                         \inst{\ref{barcelona}} 
 \and A.~Guerrier\inst{\ref{cnes}}
      \and P.~Panuzzo\inst{\ref{gepi}}
      \and D.~Katz\inst{\ref{gepi}}
      \and G. M.~Seabroke\inst{\ref{mssl}}
      \and F.~Th\'{e}venin\inst{\ref{oca}}
       %CU6 management old
      \and M.~Cropper\inst{\ref{mssl}}
      %DU leads
      \and K.~Benson\inst{\ref{mssl}}
       \and R.~Blomme\inst{\ref{brussels}}
      \and  R.~Haigron\inst{\ref{gepi}}
      \and M.~Smith\inst{\ref{mssl}}
      % active
       \and S.~Baker\inst{\ref{mssl}}
      \and L.~Chemin\inst{\ref{chile}}
       \and M.~David\inst{\ref{antwerpen}}
      \and C.~Dolding\inst{\ref{mssl}}
       \and Y.~Fr\'{e}mat\inst{\ref{brussels}}
       \and K.~Jan{\ss}en\inst{\ref{aip}}
       \and G.~Jasniewicz\inst{\ref{montpellier}}
       \and A.~Lobel\inst{\ref{brussels}}
      \and G.~Plum\inst{\ref{gepi}}
      \and N.~Samaras\inst{\ref{brussels},\ref{prague}}
      \and O.~Snaith\inst{\ref{gepi}}
      \and C.~Soubiran\inst{\ref{bordeaux}}
      \and O.~Vanel\inst{\ref{gepi}}
      \and T.~Zwitter\inst{\ref{ljubljana}}
                 %less contribution/DPCC operations/old
     \and N.~Brouillet\inst{\ref{bordeaux}}
      \and E.~Caffau\inst{\ref{gepi}}
      \and F.~Crifo\inst{\ref{gepi}}
       \and C.~Fabre\inst{\ref{atos},\ref{cnes}}
       \and F.~Frakgoudi\inst{\ref{garching}}
        \and A.~Jean-Antoine Piccolo\inst{\ref{cnes}}
      \and H.E.~Huckle\inst{\ref{mssl}}
      \and Y.~Lasne\inst{\ref{thales},\ref{cnes}}
      \and N.~Leclerc\inst{\ref{gepi}}
      \and A.~Mastrobuono-Battisti\inst{\ref{gepi},\ref{lund}}
      \and F.~Royer\inst{\ref{gepi}}
      \and Y.~Viala\inst{\ref{gepi}}
        \and J.~Zorec\inst{\ref{brussels}}
         % \textbf{\color{red} co-authors, affiliations and acknoledgements to be confirmed}
   %  \and the CU6  \inst{\ref{inst:0001}} 
}     
\institute{
GEPI, Observatoire de Paris, Universit\'e PSL, CNRS, 5 Place Jules Janssen, 92190 Meudon, France
\relax\label{gepi}
\and 
Observatoire Astronomique de Strasbourg, Universit\'{e} de Strasbourg, CNRS, UMR 7550, 11 rue de l'Universit\'{e}, 67000 Strasbourg, France                                                    
 \label{strasbourg}
 \and
Univ. Grenoble Alpes, CNRS, IPAG, 38000 Grenoble, France
\label{grenoble}
\and 
Institut de Ci\`{e}ncies del Cosmos, Universitat  de  Barcelona  (IEEC-UB), Mart\'{i} i  Franqu\`{e}s  1, 08028 Barcelona, Spain                                                                
\label{barcelona}
\and 
CNES Centre Spatial de Toulouse, 18 avenue Edouard Belin, 31401 Toulouse Cedex 9, France
\relax\label{cnes}
\and 
Mullard Space Science Laboratory, University College London, Holmbury St Mary, Dorking, Surrey RH5 6NT, United Kingdom
\label{mssl}
\and
Universit\'{e} C\^{o}te d'Azur, Observatoire de la C\^{o}te d'Azur, CNRS, Lagrange UMR 7293, CS 34229, 06304, Nice Cedex 4, France
\label{oca}
\and 
Royal Observatory of Belgium, Ringlaan 3, 1180 Brussels, Belgium
\label{brussels}
\and 
CRAAG - Centre de Recherche en Astronomie, Astrophysique et G\'{e}ophysique, Route de l'Observatoire Bp 63 Bouzareah 16340, Alger, Alg\'erie
\relax\label{alger}
\and 
Institut d'Astrophysique et de G\'{e}ophysique, Universit\'{e} de Li\`{e}ge, 19c, All\'{e}e du 6 Ao\^{u}t, B-4000 Li\`{e}ge, Belgium
\label{liege}
\and
Unidad de Astronom\'ia, Fac. Cs. B\'asicas, Universidad de Antofagasta, Avda. U. de Antofagasta 02800, Antofagasta, Chile
\label{chile}
\and 
Universiteit Antwerpen, Onderzoeksgroep Toegepaste Wiskunde, Middelheimlaan 1, 2020 Antwerpen, Belgium
\relax\label{antwerpen}
\and
F.R.S.-FNRS, Rue d'Egmont 5, 1000 Brussels, Belgium
\label{fnrs}
\and
Leibniz Institute for Astrophysics Potsdam (AIP), An der Sternwarte 16, 14482 Potsdam, Germany
\label{aip}
\and 
Laboratoire Univers et Particules de Montpellier, Universit\'{e} Montpellier, CNRS, Place Eug\`{e}ne Bataillon, CC72, 34095 Montpellier Cedex 05, France
\relax\label{montpellier}
\and Charles University, Faculty of Mathematics and Physics, Astronomical Institute of Charles University, V Hole\v{s}ovi\v{c}k\'{a}ch 2, 180 00 Prague, Czech Republic
        \label{prague}
\and Laboratoire d'astrophysique de Bordeaux, Universit\'{e} de Bordeaux, CNRS, B18N, all{\'e}e Geoffroy Saint-Hilaire, 33615 Pessac, France
\label{bordeaux}
\and
Faculty of Mathematics and Physics, University of Ljubljana, Jadranska ulica 19, 1000 Ljubljana, Slovenia
\label{ljubljana} 
\and
ATOS for CNES Centre Spatial de Toulouse, 18 avenue Edouard Belin, 31401 Toulouse Cedex 9, France
\label{atos}
\and
Max Planck Institute for Extraterrestrial Physics, High Energy Group, Gie{\ss}enbachstra{\ss}e, 85741 Garching, Germany 
\label{garching}
\and
Thales Services for CNES Centre Spatial de Toulouse, 18 avenue Edouard Belin, 31401 Toulouse Cedex 9, France
\label{thales}
\and Department of Astronomy and Theoretical Physics, Lund Observatory, Box 43, SE--221 00, Lund, Sweden
        \label{lund}
}

\date{Received March 23, 2022; accepted May 10, 2022}

\abstract
  % context
  {{\it Gaia} Data Release 3 (DR3) contains the first release of  magnitudes estimated from the integration of Radial Velocity Spectrometer (RVS) spectra for a sample of about 32.2 million stars brighter than $\grvs \sim14$~mag (or $\g\sim15$~mag).} 
   %Aims 
  {In this paper, we describe the data used and the approach adopted to derive and validate the \grvs\  magnitudes published in DR3.  We also provide estimates of the \grvs\ passband and associated \grvs\ zero-point.}
  % Methods 
  {We derived \grvs\ photometry from the integration of RVS spectra over the wavelength range from 846 to 870~nm. We processed these spectra following a procedure similar to that used for DR2, but incorporating several improvements that allow a better estimation of \grvs. These improvements pertain to the stray-light background estimation, the line spread function calibration, and the detection of spectra contaminated by nearby relatively bright sources. We calibrated the \grvs\ zero-point every 30 hours based on the reference magnitudes of constant stars from the Hipparcos catalogue, and used them to transform the integrated flux of the cleaned and calibrated spectra into epoch magnitudes. The \grvs\ magnitude of a star published in DR3 is the median of the epoch magnitudes for that star. We estimated the  \grvs\ passband by comparing the RVS spectra of 108 bright stars with their flux-calibrated spectra from external spectrophotometric libraries. }
    %  Results 
  { The \grvs\  magnitude provides information that is complementary to that obtained from the \g, \gbp, and \grp\ magnitudes, which is useful for constraining stellar metallicity and interstellar extinction. The median precision of \grvs\ measurements ranges from about 0.006~mag for the brighter stars (i.e. with $3.5\simlt\grvs\simlt6.5$~mag) to 0.125~mag at the faint end. The derived \grvs\ passband shows that the effective transmittance of the RVS is approximately 1.23 times better than the pre-launch estimate.}
  % Conclusions - Optional, can be empty if not applicable.
  {}
  

\maketitle

%%%%%%%%%%%%%%%%%%%%%%%%%%%
%                         %
% Section 1. Introduction %
%                         %
%%%%%%%%%%%%%%%%%%%%%%%%%%%

\section{Introduction\label{sec:intro}}

The high-resolution spectra collected by the Radial Velocity Spectrometer (RVS) on board \gaia\ \citep{DR1-DPACP-18} offer the possibility to define a narrow-band Vega-system \grvs magnitude linked to the effective spectral transmittance of the instrument. {\it Gaia} Data Release 3 \citep[DR3,][]{DR3-DPACP-185} contains the first release of \grvs\ magnitudes estimated by the RVS pipeline 
using the flux integrated in the RVS spectra. These magnitudes are provided for about 32.2 million bright stars 
observed by \gaia, along with the \g, \gbp, and \grp magnitudes obtained from
the astrometric images, and the blue and red photometers, respectively \citep{2021A&A...649A...3R}. 

The \grvs\ measurements are published in the 
\gaia\ archive {\tt gaia\_source} table in the column labelled {\tt grvs\_mag}. The associated uncertainties are listed in the column {\tt grvs\_mag\_error}
and the number of epoch measurements used to obtain {\tt grvs\_mag} in the column {\tt grvs\_mag\_nbtransits}.
In the following, we refer to these quantities using their \gaia\ archive name, while we use the term \extgrvs\ to designate
 \grvs\ magnitudes obtained from external, that is non-RVS, data (including rough estimates by the onboard software, \obgrvs, and finer
 estimates from the ground processing of \gaia\ astrometric images and red-photometer spectra).

The RVS was designed and optimised 
to obtain the spectra of the brightest stars observed by \gaia\ (i.e. brighter than the RVS magnitude limit of $\obgrvs\sim16.2$~mag).
The primary goal of the RVS pipeline is to measure
the all-epoch-combined radial velocities of these stars, with the measurement of {\tt grvs\_mag} being a secondary task. 
Each intermediate data release allows us to progress on these tasks,
because the processing of more epoch data implies a higher signal-to-noise ratio (S/N) of the combined RVS spectra and the possibility to 
reach fainter magnitudes. In DR3, radial velocities are provided down to ${\tt grvs\_mag}\sim14$~mag, 
compared to only $\sim12$~mag in DR2. We
aim to approach the RVS magnitude limit of $\sim16$ in {\it Gaia} Data Release 4 (DR4).

This paper, focused on {\tt grvs\_mag}, is part of a series of papers dedicated to specific products of
the RVS pipeline: the mean radial velocities are described in \citet{DR3-DPACP-159} and, for hot stars, in \citet{DR3-DPACP-151};
the double-lined radial velocities in \citet{DR3-DPACP-161}; the mean projected rotational velocities in \citet{DR3-DPACP-149}; 
and the mean spectra in \citet{DR3-DPACP-154}.

In this paper, we describe the reduction process and method used to convert the raw RVS spectra into the {\tt grvs\_mag} magnitudes
published in DR3. The paper is organised as follows. In Sect.~\ref{sec:data}, we present the RVS data used. The processing of the RVS spectra
and the estimation of  {\tt grvs\_mag} from these spectra are described in Sects.~\ref{sec:dataproc} 
and~\ref{sec:estimateRVSgrvs}, respectively. Section~\ref{sec:estimateCU5grvs} presents an alternative estimation of \grvs\ based on 
the \g and \grp magnitudes. In Sect.~\ref{sec:grvsval}, we describe the validation of {\tt grvs\_mag} and the performances achieved,
while in Sect.~\ref{sec:grvsFilter}, we compute and provide the \grvs passband.
Section~\ref{sec:grvsUse} illustrates the potential of {\tt grvs\_mag} to constrain interstellar extinction and stellar metallicity and to separate cool dwarfs from cool giants. 
Our conclusions are summarised in Sect.~\ref{sec:conc}.

%%%%%%%%%%%%%%%%%%%%%%%%%%%%%%%%%%%%%%%%%%%%%%%%
%%%%%%%%%%%%%%%%%%%%%%%%%%%%%%%%%%%%%%%%%%%%%%%
%                                             %
% Section 2. The data: describe the data used %
%                                             %
%%%%%%%%%%%%%%%%%%%%%%%%%%%%%%%%%%%%%%%%%%%%%%%

\section{The data used\label{sec:data}}

%The estimation of \grvs\ is obtained using the spectra acquired by the RVS during the 34 months covered by \gdr, and processed by the  DR3 version of the Spectroscopic Pipeline.  
 \subsection{The RVS spectra \label{sec:rvsspectra}}
We refer to \citet{DR2-DPACP-46} and \citet{DR2-DPACP-47} for a complete description of the acquisition and processing of the RVS spectra. Briefly, as \gaia\ continuously scans the sky, stars brighter than $\obgrvs=16.2$~mag\footnote{The \gaia\ onboard magnitude, \obgrvs, is obtained by the onboard software from red-photometer spectra when these are not saturated, and otherwise from the astrometric images. } transiting through one of the four rows splitting the RVS focal plane in the across-scan (AC) direction will have their spectrum recorded in each of the three CCDs along that row \citep[see figure~1 of][]{DR2-DPACP-47}, so long as the onboard limit to the number of spectra that can be obtained simultaneously is not reached \citep{DR2-DPACP-46}. This limit is set by the maximum number of 72 samples that can be read by the serial register (in the AC direction) and corresponds to a maximum density of 35 000 sources/deg$^2$. When this limit is reached, priority is given to the brighter sources. 

\begin{figure}
  \begin{center}
  \includegraphics[width=8.5cm]{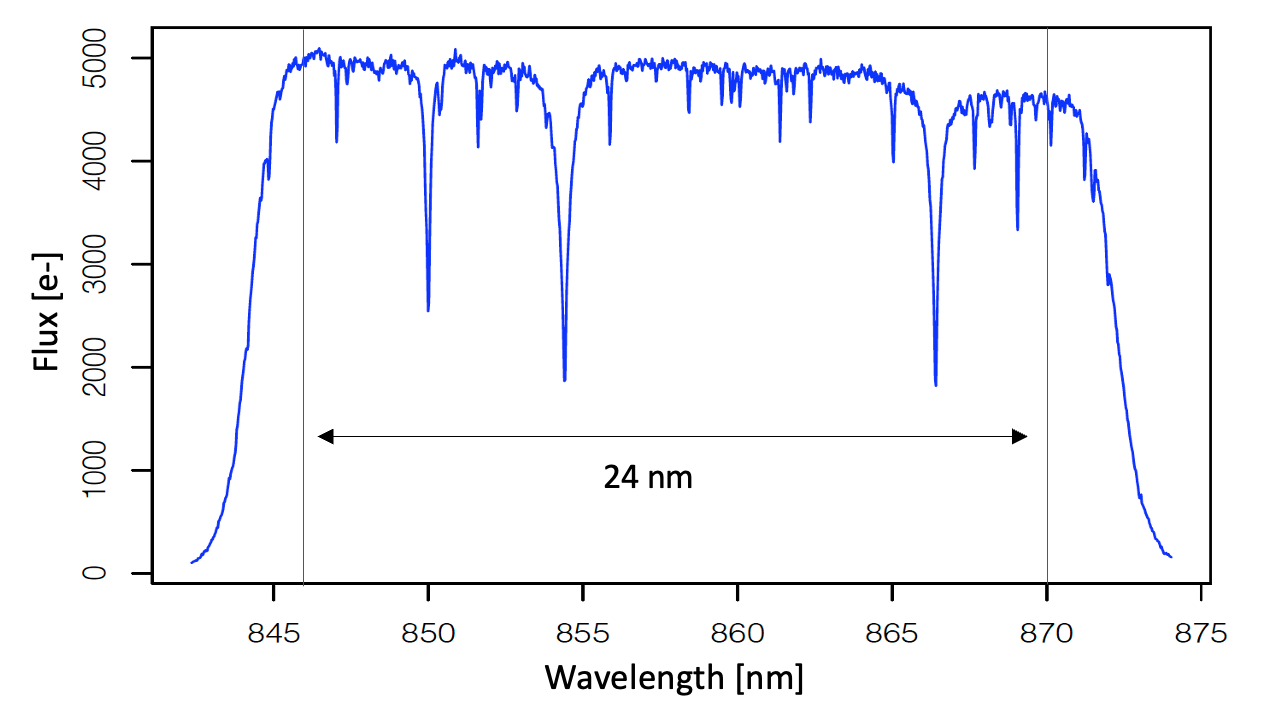}
 \end{center}
 \caption{ \grvs\ bandwidth (delimited by the two vertical orange lines) extending over 24~nm from 846 to 870~nm. The fluxes of the available RVS epoch spectra of a source are integrated between these two wavelengths to estimate the magnitude {\tt grvs\_mag}, following the procedure outlined in Sects.~\ref{sec:dataproc}~and~\ref{sec:estimateRVSgrvs}. The Ca II triplet (at 850.035, 854.444, and 866.452~nm; rest-wavelengths in vacuum), prominent in the spectra of medium-temperature, FGK-, and late M-type stars, is the dominant feature found in the majority of RVS stellar spectra. The epoch CCD spectrum shown is that of a solar-type star with  ${\tt grvs\_mag}=4.680\pm0.045$~mag.}
 \label{fig:spectrum}
 \end{figure}

In the RVS spectra, starlight is dispersed over about 1100 pixels in the along-scan (AL) direction, sampling the wavelength range from 845 to 872~nm with a resolving power of $R\approx11,500$ (resolution element  of about 3 pixels). The wings of the spectra are excluded during processing, reducing the effective length of the spectra to [846; 870] as illustrated in Fig. \ref{fig:spectrum}.

The exposure time on each of the three CCDs along a row of the RVS focal plane is fixed at 4.4 seconds by the scanning requirements, resulting in low S/Ns in the spectra of the fainter stars. As an example, the typical S/N per sample\footnote{One sample corresponds to 1\,AL\,$\times$\,10\,AC pixels.} in the spectrum of a faint source with {\tt grvs\_mag}$\,\sim14$ recorded by one of the RVS CCDs for nominal stray-light level (70 e-/sample) is only $\sim$0.7. 

 \subsection{Time gaps in the data\label{sec:baddata}}
The spectra processed by the DR3 pipeline were acquired
by the RVS between onboard mission timeline (OBMT) 1078.3795 (25 July 2014)
and OBMT 5230.0880 (28 May 2017).
The OBMT, generated by the \gaia\ onboard clock, counts the number of six-hour
spacecraft revolutions since launch. 
The relation to convert OBMT into barycentric coordinate time
(TCB) is provided by Eq.~1 of  \citet{EDR3-DPACP-130}. All events on board are given in OBMT. 

This 34-months time interval contains gaps over which the collected data were of poor quality and 
could not be used by the RVS pipeline. These gaps, when added together, cover 7.8\% of the total observing time
and were mostly caused by spacecraft events. The largest gaps were caused by three
decontamination campaigns starting at OBMT 1317, 2330, and 4112.8, each requiring about 70 revolutions
for the satellite to reach thermal equilibrium again.
The start and end times of the gaps used by the RVS pipeline are available from the cosmos pages.
 \footnote{ https://www.cosmos.esa.int/web/gaia/dr3-data-gaps}

 \subsection{Source selection: $\extgrvs\le14$ \label{sec:selectextgrvs}}

{\tt grvs\_mag} cannot be used to preselect sources to be processed by the RVS pipeline because it is a final product of this pipeline.
A \extgrvs magnitude, measured from non-RVS data (Sect.~\ref{sec:intro}), must be used instead.
 
Neither the \g\ magnitude (measured from the \gaia\ astrometric images) nor the \grp\ magnitude (measured 
from the red-photometer data) published in \gaia-EDR3 \citep{EDR3-DPACP-130} were available at the time the DR3 RVS processing started. 
Thus, estimates of these quantities from DR2 data were adopted instead to compute \extgrvs using the 
transformation formulae in Eqs.~(2) and (3) of \citet{DR2-DPACP-36}. 
This could not be achieved for new sources observed since DR2 or sources for which the source identifier had changed.
For such sources, the onboard magnitude \obgrvs was used to estimate \extgrvs. These sources
represent about 2.6\% of our final sample.

The sources with $\extgrvs\le14$ from this sample represent about 20\% of the spectra observed by the RVS 
over the period of interest (Sect.~\ref{sec:baddata}). The selection on the basis of \extgrvs also implies that
some sources making the magnitude cut can end up with a {\tt grvs\_mag} measurements of fainter than 14 mag. 
Measurements of {\tt grvs\_mag} of fainter than 14.1 mag were considered spurious, 
mostly affected by inaccurate background estimation,
and have not been published (see Sect.~\ref{sec:filters}).

 \subsection{Spectrum selection\label{sec:epoch selection}}
  
 For the purpose of {\tt grvs\_mag} estimation, only the clean spectra are selected. To limit
 the contamination from nearby sources, the spectra presenting a truncated window (i.e.
 for which the spectrum window on the CCD overlaps with that of
 a nearby source) are excluded (some of these spectra are still used for the radial velocity estimation, after having
 been deblended; see
 \citealt{EDR3-DPACP-121} and \citealt{DR3-DPACP-154}). 
 
 Other spectra excluded during the processing (see Sect.~\ref{sec:dataproc}) are: 
 the spectra with non-truncated window,
 but still potentially contaminated by nearby relatively bright sources;
 the spectra acquired over bad pixels or in a region with overly high levels of stray light; 
 those with a negative total flux after bias and background subtraction; and
 those with too many samples saturated or affected by cosmic rays. In the end, of the 2.8 billion spectra with $\extgrvs\le14$ treated by the DR3 RVS pipeline (100 times more spectra than for DR2), 
about 1.5 billion are retained to estimate {\tt grvs\_mag}.

%%%%%%%%%%%%%%%%%%%%%%%%%%%%%%%%%%%%%%%%%
%                                       %
% Section 3. RVS spectra cleaning and calibration %
%                                       %
%%%%%%%%%%%%%%%%%%%%%%%%%%%%%%%%%%%%%%%%%

\section{Processing of RVS spectra}\label{sec:dataproc}

The updated DR3 pipeline is described in
the online documentation \footnote{\url{https://gea.esac.esa.int/archive/documentation/GDR3/Data_processing/chap_cu6spe/ }} \citep{CU6-DR3-documentation}, and the algorithms for performing the cleaning and calibration of the RVS spectra are described in detail in \citet[][Sects~5 and 6]{DR2-DPACP-47}.  In this section, we summarise the processing steps relevant to the computation of {\tt grvs\_mag}. We also describe in some detail new functionalities of the pipeline that allowed improvements in the estimation of grvs\_mag, such as
stray-light background estimation and the identification of spectra contaminated by neighbouring spectra.

\subsection{The processing steps}\label{sec:procsteps}

Each individual RVS CCD spectrum passes through the following processing steps relevant to {\tt grvs\_mag} estimation.
The aim is to compute {\it TotFlux}, the star flux integrated from the RVS spectrum.
%{\bf please remove the bullet point format and supply these points, numbered, in paragraph form in continuous prose.}

1. The flux in the raw spectrum, in  Analog Digital Units (ADU), is corrected for electronic bias and non-uniform offset \citep{DR2-DPACP-29}.

\begin{figure}[!ht]
  \begin{center}
  \includegraphics[width=8cm]{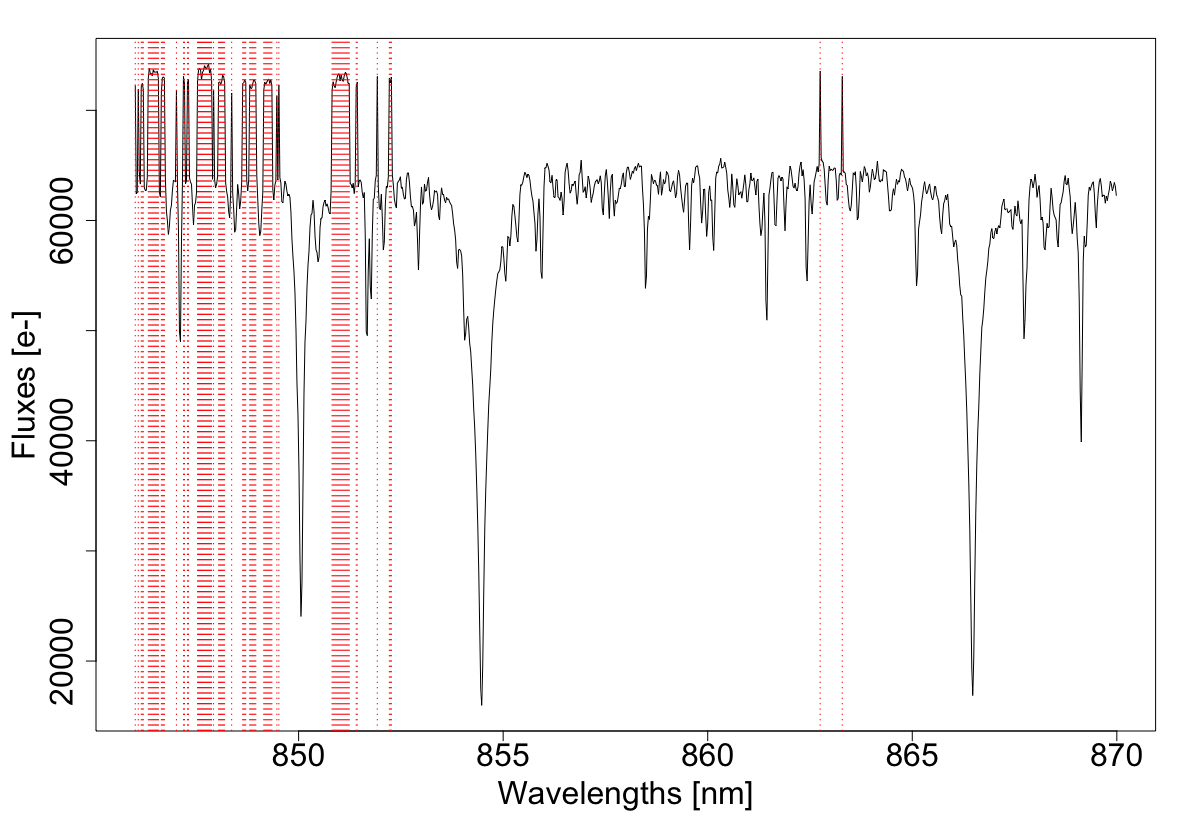}
 \end{center}
 \caption{Saturation in an RVS spectrum. The CCD spectrum shown (in units of e$^-$ per 4.4~sec exposure) was obtained from the summing in the AC direction of a 2D-window spectrum including saturated AC-central pixels. The resulting 1D spectrum, which presents 99 saturated samples (marked in red), was discarded by the pipeline. The visible jumps (of $\sim8000\,\mathrm{e}^-$) originate from AC-central pixels exceeding 50 000~ADU (see Sect.~\ref{sec:procsteps}). The corresponding star has ${\tt grvs\_mag}=3.471\pm0.005$~mag, $T_{\rm eff}=4735$~K, $\log g=1.45$, and ${\rm [Fe/H]}=-0.13$.}
 \label{fig:saturatedspe}
 \end{figure}

2. All pixels with fluxes exceeding $\sim$ 50 000~ADU are flagged as `saturated'. In practice, such pixels are assigned the numerical-saturation value of 65 535 ADU, as shown in Fig.~\ref{fig:saturatedspe}. This procedure flags by default all pixels reaching physical saturation, because the average full-well capacity of an RVS CCD pixel is 190 000~e$^-$, corresponding to 336 300 ADU \cite[see Table 1 of][]{DR2-DPACP-29}. Spectra presenting saturated pixels are not used for the \grvs\ zero-point estimation (Sect.~\ref{sec:grvszp}), while a maximum of 40 saturated pixels are allowed for the estimation of the \epochgrvs\ epoch magnitude (Sect.~\ref{sec:grvsmag}).

3. The flux is transformed into photoelectrons using the on-ground-measured CCD gain values (0.53 e$^-$ ADU$^{-1}$).

4. The stray-light background is subtracted from the spectrum (see Sect.~\ref{sec:background} for more details). The spectra with negative total flux caused by over-subtraction of the background are removed from the pipeline, which induces a systematic overestimate of the flux of faint stars (see Sect.~\ref{sec:grvssimu}). 
Also removed are the spectra for which the background is too high (i.e. higher than 100~e$^-$\,pixel$^{-1}$\,s$^{-1}$; or higher than 40~e$^-$\,pixel$^{-1}$\,s$^{-1}$ with an uncertainty higher than 0.4~e$^-$\,pixel$^{-1}$\,s$^{-1}$).

5. The flux loss outside the spectrum window on the CCD is estimated using the model of line spread function in the AC direction (LSF-AC) obtained in the pipeline. The LSF-AC calibration is a new functionality of the DR3 pipeline and is described in the online documentation \citep[][Sect. 6.3.4]{CU6-DR3-documentation}.  The LSF-AC profile is measured over an AC pixel range of  $\pm(5+2.5)$ pixels from the centre (i.e. out to 2.5 pixels on each side beyond the 10-pixel-wide window in the AC direction). Outside of this range, the AC LSF is extrapolated to zero at $\pm$20 pixels. The flux loss outside the window is estimated using the extrapolated LSF-AC profile (i.e. over 15 pixels, or 2.67\arcsec, on each side of the nominal window) and is typically of the order of 5\%. 

6. Spectra containing any column from the cosmetic-defect list (see the online documentation: \citealt{CU6-DR3-documentation}, table~6.2) are flagged and removed from the pipeline.

7. Spectra contaminated by a nearby source are flagged and removed from the pipeline (see Sect.~\ref{sec:contam} below for a description of the detection of contaminants).

8. Cosmic rays are removed. If the number of pixels affected by cosmic rays reaches 100, the spectrum is removed from the pipeline.

9.  2D windows \citep[pertaining to stars with $\obgrvs\leq7$; see figure~1 of][]{DR2-DPACP-47} are optimally collapsed \citep{1986PASP...98..609H} into 1D spectra if there are no saturated pixels. Otherwise, the 2D windows are collapsed into 1D spectra with a simple summing in the AC direction. 

10. The wavelength calibration is applied, and the wavelength range is cut to 846--870~nm to remove the wings of the RVS spectrum (Fig.~\ref{fig:spectrum}). This is the widest possible (integer) wavelength range properly sampled by all spectra of a given source, as the spectra obtained over various transits are not uniformly sampled in the wings. 
%{\bf A and A insists footnotes be incorporated into the body text when discursive and where possible.Please incorporate this footnote into the text}\footnote{
In fact, each RVS observation window is divided into 12 subunits of 108 AL pixels, called macrosamples \citep{DR2-DPACP-46}. To limit the processing load on board, all windows in a given CCD are phased at macrosample level and can start only at macrosample boundaries. As the centring of a spectrum in the window depends on the observing configuration, the spectra of a given source obtained in different observation windows may have their ends cut off by up to 108 AL pixels, implying non-uniform sampling in the extreme macrosamples.
%} 
We also note that, as the LSF-AL profile \citep[][Sect. 6.3.4]{CU6-DR3-documentation} contributing to the wings of the spectra differs from one position to another in the focal plane, the cutoff for a given star will correspond to a different fraction of flux lost depending on the position of the epoch observation. This effect is included in the standard deviation of the measurements.

11.  In the bright spectra (\extgrvs\ $\le 12$), emission lines (whether real or spurious) are detected and flagged (affected spectra are not used for the \grvs\ zero-point calibration; see Sect. \ref{sec:grvszp}).

12.  In the bright spectra  (\extgrvs\ $\le 12$), the presence of flux gradients is detected by comparing the median fluxes at the blue (between 846 and 849~nm ) and red (between 867 and 870~nm) edges. If the ratio between the red- and blue-edge fluxes is greater than 1.2, the spectrum is flagged as having a positive gradient. If it is less than 1/1.2, it is flagged with negative gradient. The presence of a gradient may indicate potential problems (such as mis-centring of the acquisition window on the source, a data processing issue, etc.), although in cool stars, a positive gradient may simply indicate the presence of TiO molecular absorption. For bright stars hotter than 3500~K, about 0.4\% of all spectra are flagged with positive gradients, and about 0.7\% with negative gradients. 

13.  After replacing the flux in any flagged sample (saturated or affected by cosmic rays) with the median flux over all good samples, the total flux of the spectrum in the \grvs window, {\it TotFlux}, is estimated by summing the fluxes of all samples between 846 and 870 nm. Then, {\it TotFlux} is corrected for the estimated flux loss outside the window and divided by the 4.4~s exposure time to be expressed in units of \es.

%%%%%%%%%%%n%%%%%%%%%%%%%%%%%%
%%%%%%%%%%%%%%%%%%%%%%%%%%%%%%

\subsection{Estimation of the stray-light background}\label{sec:background}
The RVS spectra are contaminated by a diffuse background dominated by solar stray light caused by diffraction from detached fibres in the sunshield (Cropper et al. 2018). 
While stray-light contamination varies over time and also with position in the focal plane, it follows, for the most part, a relatively stable pattern related to the satellite rotation phase. Crowding effects also contribute to the background estimation: in crowded regions, such as during Galactic plane scans, the level of diffuse light contributing to the background is higher.

The accuracy and precision of the background measurement impact estimates of the total flux in the individual RVS spectra. The precision of the background estimation was improved in the DR3 pipeline through regular calibrations and the use of information outside the filter passband (i.e. at wavelengths below 843.2~nm and above 874.2~nm) in faint-star spectra to increase the number of individual background measurements.

Specifically, the background level was estimated every five spacecraft revolutions (i.e. every 30\,hr). On each occurrence, the background was measured using the fluxes of so-called virtual objects (VOs; corresponding to empty windows with only background signal), together with flux measurements at the outer edges of the spectral windows of stars fainter than $\obgrvs=15$\,mag. The background associated with each RVS spectrum is the median flux computed using all clean VOs and faint stars in a large area around the star of interest, over a period of $\pm6$ revolutions ($\pm36$\,hr) around the time of estimation. The area corresponds to 36 seconds of scan (per 6\,hr revolution) including the star position, which samples 36 600 pixels in the AL direction (i.e. $\sim2160\arcsec$) and 251 pixels in the AC direction ($\sim45\arcsec$). We note that this procedure smoothes out any local variation of the background level over the considered area, as well as any temporal variation over a period of 72 hr (see section 5.2 of \citealt{DR2-DPACP-47} for a detailed description of the background estimation algorithm). 
 
The standard deviation of residuals around the median background estimate amounts to typically $3.15$\,e$^-$\,sample$^{-1}$  in a 4.4 s exposure. This yields a typical uncertainty of $\pm$702 \es\ in the estimated background flux integrated over an RVS spectrum (about 980 samples).  Hence, the background flux uncertainty for a single RVS-CCD spectrum is similar to the expected flux of a star with ${\tt grvs\_mag}=14.1$\,mag (for which the flux uncertainty would therefore be of 100\%).

%%%%%%%%%%%%%%%%%%%%%%%%%%%
%%%%%%%%%%%%%%%%%%%%%%%%%%%

\subsection{Contamination from nearby sources}\label{sec:contam}

Some light from nearby sources may enter the RVS-spectrum window of the target source, leading to overestimation of the flux. Here, `nearby'  is meant in the focal-plane reference system: the two \gaia\ telescopes share the same focal plane \citep{DR1-DPACP-18}, and nearby sources in the focal plane may come from the two different fields of view (FoVs) and be physically very far apart.

The relative distance between the contaminant and the target source in the focal plane is epoch-dependent. An RVS window may or may not have been assigned to the contaminant during a specific satellite scan. This is because of the onboard limit on the number of RVS spectra that can be obtained simultaneously, and occurs when one of the \gaia\ FoVs scans a crowded region of the sky. We define as `contamination area' the area in the focal plane centred on the target source that extends over twice the size of the regular window. This contamination area corresponds to about 2592\,AL\,$\times$\,21\,AC pixels, or 152.93\arcsec\,AL\,$\times$\,3.74\arcsec\,AC. Nearby sources in the contamination area have a different impact on the target window depending on whether or not they have an RVS window. The two cases are therefore treated differently in the pipeline:

1.  Contaminants with an RVS window generate truncations in the target source window (because the two windows partly overlap). This type of contamination is naturally accounted for by not using the target spectra with truncated windows.\footnote{In rare cases, affecting faint stars in crowded areas, two windows can completely overlap, or overlap over only one or two pixels; both situations interfere with the standard pipeline processing. Such cases are described in detail in section~2.4 of \citet{DR3-DPACP-154}.} As an exception, 2D windows \citep[pertaining to very bright stars with $\obgrvs\leq7$; see figure~1 of][]{DR2-DPACP-47} are not truncated even if they are in conflict with other source windows. Contamination of 2D window spectra by relatively bright nearby sources is rare in practice and is ignored by the pipeline.
 
 2.  Contaminants without an RVS window do not generate any window conflict or window truncation. Potential contaminants are identified as transits of \gaia-catalogue sources brighter than $\extgrvs = 15$ (fainter contaminants are ignored) and without an RVS window, based on the `ObjectLogsRVS' files produced on board. The predicted AL and AC positions of the potential contaminants are computed by projecting the known astrometric coordinates of the source onto the focal plane, taking into account the satellite attitude and geometry. The effective contaminants are defined as the sources located in the contamination area around the target and which are sufficiently bright relative to the target, that is, brighter than \extgrvs(target)+3. 
%
% i.e. a contaminant brighter than 3 magnitudes fainter than the target is considered as relatively bright; a target source of magnitude 10 can be contaminated by nearby sources brighter than 13).  
%
Transits of target sources with such contaminants are removed, while those of target sources with fainter contaminants are flagged `faint-contaminated' and are used to estimate {\tt grvs\_mag}. 

A total of about 135 million CCD spectra were removed because of contaminants without an RVS window.

With this procedure, for most of the target stars fainter than $\obgrvs=7$, we expect to exclude all transits affected by contaminants closer than $\sim1\,.87\arcsec$ and, depending on the satellite scan direction, at least part of the transits affected by contaminants at distances between $\sim1.87\arcsec$ and $\sim76.46\arcsec$.

Potential contamination from relatively bright sources (with or without an RVS window) located outside the contamination area is ignored in the pipeline. Neither the RVS filter response nor the LSF-AC was calibrated significantly outside the RVS window. Based on the on-ground data and extrapolation of the LSF-AC calibration, a contaminant just outside the contamination area is expected to contaminate the target window with less than $10^{-4}$ of its flux in the AL direction, and with less than $\sim 2$\% of its flux in the AC direction (i.e. to have more than 0.2 mag of contamination on the target source, a contaminant located outside the contamination area must be at least 2.5 mag brighter than the target; very bright contaminant sources outside the contamination area can therefore still contaminate targets at the faint end). Also ignored by the pipeline are the potential contaminants with no RVS window and no \extgrvs\ estimate (having changed source identifiers since DR2), and of course, those not present in the \gaia-source catalogue at all (not observed by \gaia).
 
Finally, a validation procedure based on the filters described in Sect.~\ref{sec:filters} eliminated many spurious estimates of {\tt grvs\_mag} affected by contamination from bright nearby sources, including outside the contamination area.

%%%%%%%%%%%%%%%%%%%%%%%%%%%%%%%%%%%%%%%%%%%%%%%%%%%%%%%%%%%%%%%%%%%%%%
%                                                                    %
% Section 4. estimate GRVS from the RVS spectra%
%                                                                    %
%%%%%%%%%%%%%%%%%%%%%%%%%%%%%%%%%%%%%%%%%%%%%%%%%%%%%%%%%%%%%%%%%%%%%%

\section{{\tt grvs\_mag} estimation from the RVS spectra}\label{sec:estimateRVSgrvs}

\subsection {Zero-point calibration}\label{sec:grvszp}

The zero-point is calibrated every 30 hr in each of the 24 RVS `configurations' (corresponding to 12 CCD in two FoVs), based on the reference magnitudes of a set of calibrator stars. The calibrator stars are the 103 865 stars in the {\it Hipparcos} catalogue \citep{1997A&A...323L..49P}  with reference magnitudes brighter than $\refgrvs=10$~mag and expected to be constant from \gaia-DR2 photometric data. \refgrvs is computed using the transformation provided in \citet{2010A&A...523A..48J}:\\
\begin{equation}
 G_\mathrm{RVS}^{\mathrm{ref}} = V - 0.0501 - 1.1667(V-I) + 0.0052(V-I)^{2} + 0.0011(V-I)^{3}\,,
 \end{equation}
where $V$ and $I$ are the magnitudes in the {\it Hipparcos} catalogue.

For a given `Calibration Unit' (CaU; corresponding to 30 hr of observations in the same RVS configuration), the {\it TotFlux} of each calibrator star observed with the RVS is computed from the spectrum as described in Sect.~\ref{sec:procsteps}. This allows the zero-point to be estimated as
\begin{equation}
ZP_{\rm{spec}} = G_\mathrm{RVS}^{\mathrm{ref}}  +  2.5 \log({\it TotFlux})\,.
\end{equation}
The global CaU zero-point is taken to be the median $ZP_{\rm{spec}}$ over all exploitable calibrator spectra (typically 150 to 200),
 \begin{equation}
 \label{zp_cau}
 ZP_{\rm{CaU}} = {\rm Med}(ZP_{\rm{spec}}),
 \end{equation}
and the global associated uncertainty the robust dispersion,
$\sigma_{ZP_{\rm{CaU}} }= \frac{P(ZP_{\rm{spec}}, 84.15) - P(ZP_{\rm{spec}}, 15.85)}{2}$\,,
where $P(ZP_{\rm{spec}}, 84.15)$ and $P(ZP_{\rm{spec}}, 15.85)$ are the 84.15$^{\rm th}$ and the 15.85$^{\rm th}$ percentiles of the $ZP_{\rm{spec}}$ distribution.

\begin{figure}[!ht]
  \begin{center}
  \includegraphics[width=8.5cm]{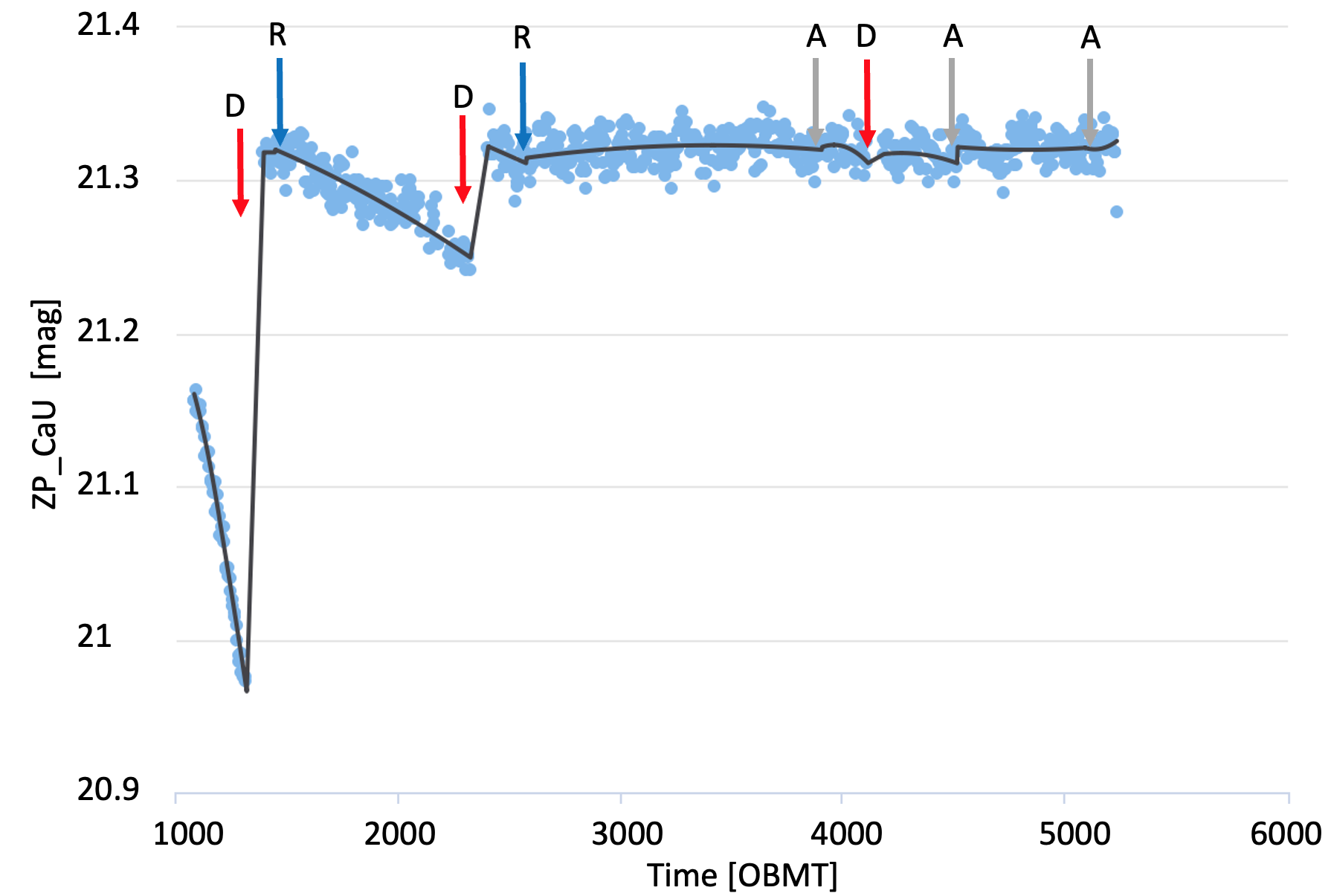}
 \end{center}
 \caption{\grvs zero-point plotted against OBMT. The blue points represent the $ZP_{\rm CaU}$ estimates (from Eq.~\ref{zp_cau}) every 30 hr for the CCD in row 6, strip 2, and FoV 2 \citep[see figure~1 of][]{DR2-DPACP-47} over the DR3 observing period. The black line is the calibration model for $ZP(t)$. The arrows indicate break points between which the calibration model is fitted. The break points correspond to the following events:  decontaminations (red arrows), refocus (blue arrows), and discontinuities in the astrometric solution (grey arrows). At the beginning of the mission, the \gaia\ optics suffered from heavy water ice contamination, which resulted in rapid degradation of the ZP. The first two decontamination events, at OBMT 1317 and 2330, produced a significant improvement in ZP.  After the second decontamination event, the ZP stabilised, and the third decontamination event, which is also the last one performed on \gaia, resulted in no significant improvement. The other events have no significant effect on the ZP. }
 \label{fig:grvszptrend}
 \end{figure}

Figure~\ref{fig:grvszptrend} shows the zero-points of all the CaUs obtained during the DR3 observing period in one of the \gaia\ CCDs. 
The time sequence of $ZP_{\rm{CaU}}$ values defines the temporal variation of the zero-point, noted $ZP(t)$, which can be modelled with second-degree polynomial trending functions, as illustrated. Generalising over all CaUs, the $ZP$ may be obtained from such functions for any RVS spectrum observed at any time and in any configuration.
%It is used to estimate the magnitude \intgrvs (Sect. \ref{sssec:intmag}). 
It is worth noting that, in the future DR4, estimates of the $ZP$ dispersion will be improved by informing the pipeline with products made available by DR3, such as reference synthetic magnitudes computed using externally calibrated low-resolution spectra from the red photometer \citep{DR3-DPACP-120} convolved with the \grvs\ passband derived in Sect.~\ref{sec:grvsFilter} below.

\subsection {{\tt grvs\_mag} estimation}\label{sec:grvsmag}

The {\tt grvs\_mag} magnitude of a star is computed on the basis of epoch magnitudes estimated each time the star is observed by the RVS.
An epoch magnitude is defined as the median\footnote{The mean is used if only two clean spectra are available.} of the three magnitudes estimated from the three CCD spectra acquired when the star is scanned by the RVS \citep[see figure~1 of][]{DR2-DPACP-47}, that is,
\begin{equation}
G^{\rm{epoch}}_\mathrm{RVS} = {\rm Med}(G^{\rm{spec}}_\mathrm{RVS})\,,
\end{equation}
where the individual magnitudes are estimated using the quantity {\it TotFlux} (Sect.~\ref{sec:procsteps}) and the zero-point at time $t_{\rm obs}$ of observation (Sect.~\ref{sec:grvszp}),
\begin{equation}
G^{\rm{spec}}_\mathrm{RVS} = - 2.5 \log({\it TotFlux}) + ZP(t_{\rm obs})\,.
\label{eq:grvsflux}
\end{equation}

The source magnitude, {\tt grvs\_mag}, is defined as the median of all epoch magnitudes,
\begin{equation}
{\tt grvs\_mag} = {\rm Med}(G^{\rm{epoch}}_\mathrm{RVS})\,.
\label{equ:rvmedian}
\end{equation}
The formal error (assuming the normal law) on this median measurement is 
\begin{equation}
\sigma_{\mathrm{Med}} = \sqrt{\frac{\pi}{2}}.\frac{\sigma( G^{\rm{epoch}}_\mathrm{RVS} )}{\sqrt{\texttt{grvs\_mag\_ntransits}}}\,,
\end{equation}
where  $\sigma(G^{\rm{epoch}}_\mathrm{RVS})$ is the standard deviation of the epoch measurements and {\tt grvs\_mag\_ntransits} is the total number of epochs. To estimate the uncertainty on {\tt grvs\_mag}, we add in quadrature to $\sigma_{\mathrm{Med}}$ an error of 0.004~mag to account for calibration-floor uncertainties:
\begin{equation}
{\tt grvs\_mag\_error} = \sqrt{\sigma_{med}^{2} + 0.004^2}\,.
\end{equation}

We note that, as already mentioned in Sect.~\ref{sec:procsteps}, our procedure to discard spectra with negative {\it TotFlux} and compute {\tt grvs\_mag} with the median of the epoch magnitudes leads to systematic underestimation of {\tt grvs\_mag} for faint stars. We attempt to quantify the resulting bias in Sect.~\ref{sec:grvssimu} below. In the future DR4, we plan to avoid this bias by computing {\tt grvs\_mag} based on the median of the flux measurements (including negative ones).

%%%%%%%%%%%%%%%%%%%%%%%%%%%%%%%%%%%%%%%%%%%%%%%%%%%%%%%%%%%%%%%%%%%%%%
%                                                                    %
% Section -  Estimate GRVS from G and RP (from EDR3)%
%                                                                    %
%%%%%%%%%%%%%%%%%%%%%%%%%%%%%%%%%%%%%%%%%%%%%%%%%%%%%%%%%%%%%%%%%%%%%%

\section{\grvs\ estimation from \g\ and \grp\ }\label{sec:estimateCU5grvs}
%%%%%%%%%%%%%%%%%%%%%

As mentioned in Sect.~\ref{sec:selectextgrvs}, 
we had  to resort to DR2 measurements of \g\ and \grp\ (the magnitudes from \gaia\ astrometric images and red photometer), together with the transformation formulae in Eqs.~(2) and (3) of \citet{DR2-DPACP-36}, to estimate \extgrvs required for pre-selecting RVS spectra for {\tt grvs\_mag} estimation. We can now use DR3 estimates of {\tt grvs\_mag}, \g,\ and \grp\ to update the formulae to estimate \grvs\ from \g\ and \grp.  We refer to this updated estimate of  \extgrvs as $G^{\rm{G,RP}}_\mathrm{RVS}$, which we compare with {\tt grvs\_mag} in Sect.~\ref{sec:accuracy} below.

\begin{figure}
  \begin{center}
\includegraphics[width=8.5cm]{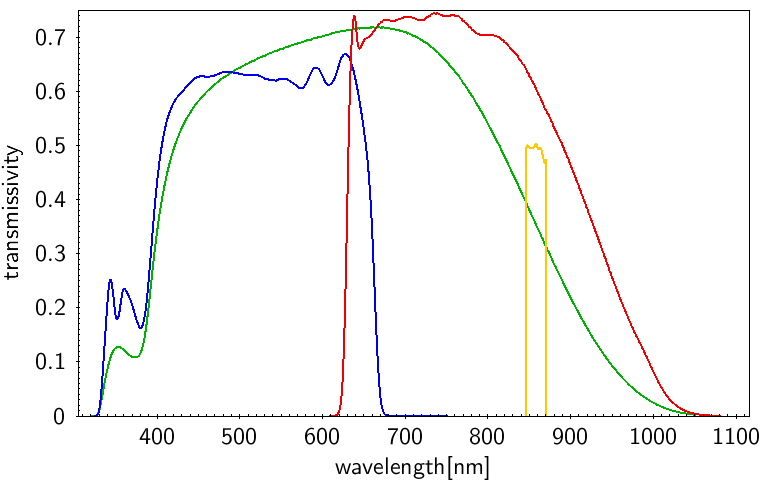}
 \end{center}
 \caption{Passbands of the \gaia\ \gbp\ (blue curve), \g\ (green curve), and \grp\ (red curve) filters \citep[from][]{2021A&A...649A...3R}. The \grvs\ passband\protect\footnotemark\ (yellow curve), which defines {\tt grvs\_mag}, is that derived in Sect.~\ref{sec:grvsFilter}.}
 \label{fig:passbands}
 \end{figure}

It is important to note that the {\tt grvs\_mag} bandwidth defined over the 846--870~nm wavelength range (Fig.~\ref{fig:spectrum}) is far narrower than the \g, \gbp, and \grp\ bandwidths of standard \gaia\ photometry (see Fig.~\ref{fig:passbands}). Colour--colour relationships between {\tt grvs\_mag}, \g,\ and \grp\ were derived from a random sample of about 3 million sources well behaved in $G$, \gbp, and \grp\ excess flux 
\citep[see][]{2021A&A...649A...3R} and with ${\tt grvs\_mag\_error}<0.05$~mag.

\footnotetext{\label{fn:pb} \url{https://www.cosmos.esa.int/web/gaia/dr3-passbands}}

\begin{figure}
  \begin{center}
\includegraphics[width=8.5cm]{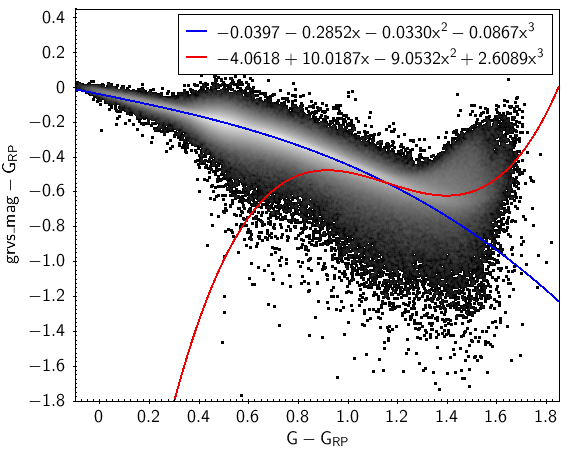}
 \end{center}
 \caption{{\tt grvs\_mag}-$\grp$ versus $\g-\grp$ colour--colour diagram for a random sample of about 3 million sources. Overplotted are the polynomial relations of eqs~(\ref{equ:bluerel}) and (\ref{equ:redrel}).}
 \label{fig:grvs_rp_vs_g_rp}
 \end{figure}

The resulting cubic polynomial fits are:

\noindent for $-0.15 \le G-G_{\rm RP} \le 1.2$,
\begin{multline}
G^{\rm{G,RP}}_\mathrm{RVS}-G_{\rm RP}=-0.0397-0.2852(G-G_{\rm RP}) \\
-0.0330(G-G_{\rm RP})^2-0.0867(G-G_{\rm RP})^3\,,
\label{equ:bluerel}
\end{multline}

\noindent
and, for $1.2<G-G_{\rm RP}\le 1.7$,
\begin{multline}
G^{\rm{G,RP}}_\mathrm{RVS}-G_{\rm RP}=-4.0618+10.0187(G-G_{\rm RP}) \\
-9.0532(G-G_{\rm RP})^2+2.6089(G-G_{\rm RP})^3\,.
\label{equ:redrel}
\end{multline}
The root mean square errors in Eqs.~(\ref{equ:bluerel}) and (\ref{equ:redrel}) are 0.04 and 0.09~mag, respectively. The two polynomial relations are overplotted on the data in the colour--colour diagram of Fig.~\ref{fig:grvs_rp_vs_g_rp}.

%%%%%%%%%%%%%%%%%%%%%%%%%%%%%%%%%%%%%%%%%%%%%%%%%%%%%
%                                                   %
% Section 6. Grvs validation         %
%                                                   %
%%%%%%%%%%%%%%%%%%%%%%%%%%%%%%%%%%%%%%%%%%%%%%%%%%%%%

%%%%%%%%%%%%%%%%%%%%%
%%% Validation
%%%%%%%%%%%%%%%%%%%%

\section{Validation of {\tt grvs\_mag}}\label{sec:grvsval}

\subsection{Underestimate of {\tt grvs\_mag} at faint magnitudes}\label{sec:grvssimu}

\begin{figure}
  \begin{center}
  \includegraphics[width=7.5cm]{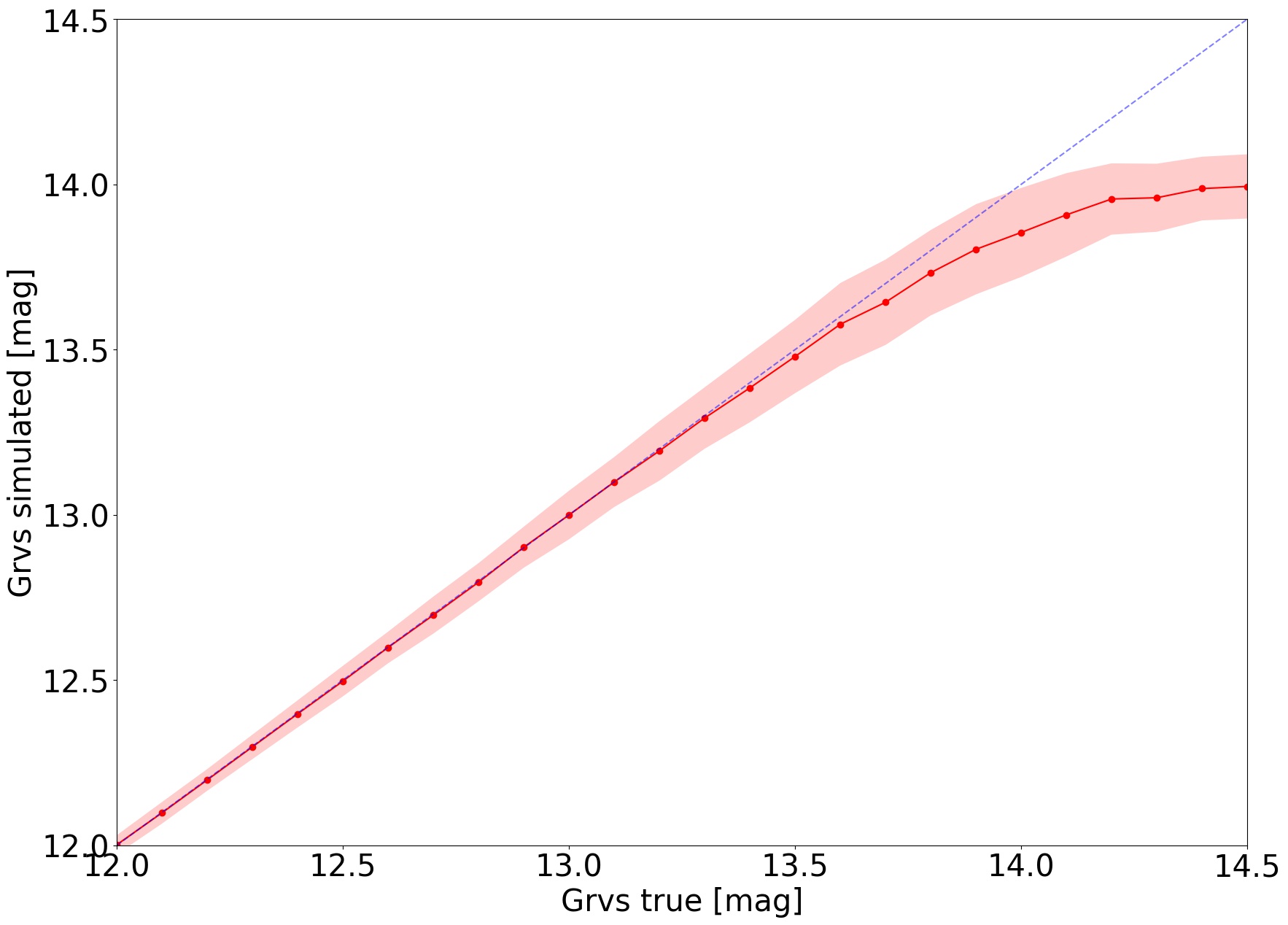}
 \end{center}
 \caption{{\tt grvs\_mag} computed following the procedure outlined in Sect.~\ref{sec:grvsmag}, $G^{simu}_\mathrm{RVS}$, plotted against true magnitude, $G^{true}_\mathrm{RVS}$, for the simulations described in Sect.~\ref{sec:grvsval}. The red line shows the median relation from 1000 realisations, and the shaded area the associated error. The dashed blue line shows the identity relation. The estimated magnitude becomes systematically brighter than the true one for $G^{true}_{\rm RVS}\simgt13.3$. The limiting magnitude of the stars processed for DR3 corresponds to $G^{true}_\mathrm{\rm RVS}=14$.}
 \label{fig:grvsfaintsimu}
 \end{figure}

We now quantify how our removal of RVS spectra with negative total flux (${\it TotFlux} \le 0$) in Sect.~\ref{sec:grvsmag} impacts our measurements of {\tt grvs\_mag}. We achieve this 
by estimating {\tt grvs\_mag} for a set of simulated stars with known true magnitudes, noted $G^{true}_\mathrm{RVS}$, between 12 and 14.5~mag (in steps of 0.1~mag.). For each star, of true flux $10^{-0.4 (G^{true}_\mathrm{RVS}- ZP)}$ (with $ZP=21.317$; see Sect.~\ref{sec:grvsFilter}), we perform 39 realisations of the total RVS flux {\it TotFlux} assuming a normal distribution of the background subtraction uncertainty centred on 702 \es. Here, 39 is the median number of individual RVS CCD spectra available to estimate {\tt grvs\_mag} for the sources in our sample (13 observation epochs $\times$ 3 CCDs). As in the procedure described in Sect.~\ref{sec:grvsmag}, we remove negative fluxes and compute a simulated  {\tt grvs\_mag}, noted $G^{simu}_\mathrm{RVS}$, using Eq.~(\ref{equ:rvmedian}). We produce 1000 such estimates of $G^{simu}_\mathrm{RVS}$ for each $G^{true}_\mathrm{RVS}$. In Fig.~\ref{fig:grvsfaintsimu}, we show the resulting median $G^{simu}_\mathrm{RVS}$ (along with its associated error) against $G^{true}_\mathrm{RVS}$. The median simulated {\tt grvs\_mag} starts to deviate from the true magnitude and become systematically brighter (by about 0.015~mag) at $G^{true}_\mathrm{RVS}=13.4$, reaching an offset of $\sim0.14$~mag at $G^{true}_\mathrm{RVS}$=14.0. A similar magnitude term is reported in \citet{DR3-Car}.

%%%%%%%%%%%%%%%%%%%%%%%%%%%%%%
%%%%%%%%%%%%%%%%%%%%%%%%%%%%%%
\subsection{Validation filtering of grvs\_mag}\label{sec:filters}

After completion of the DR3 processing of RVS data, a validation campaign was led to identify potentially erroneous data and filter these out from publication (in practice, the deemed erroneous data are nullified). Of the $\sim$37.1 million {\tt grvs\_mag} measurements produced by the pipeline, $\sim$4.9 million were nullified in this way, leading to the publication of about 32.2 million {\tt grvs\_mag}  values in DR3. In the following paragraphs, we describe the filters applied to validate these measurements.
%37,134,742 ---  4,902,308

As mentioned in Sect.~\ref{sec:intro}, the primary product of the RVS pipeline is the {\tt radial\_velocity}, and all filters applied to nullify spurious radial velocity measurements were applied to all pipeline products, including {\tt grvs\_mag}. The {\tt radial\_velocity} filter criteria are listed in the online documentation \citep[section 6.5.2.1]{CU6-DR3-documentation} and described in detail in \citet{DR3-DPACP-159} and  \citet{DR3-Car}. Most {\tt radial\_velocity} filters were beneficial also for {\tt grvs\_mag}, leading to the removal of many spurious  {\tt grvs\_mag} measurements contaminated by nearby relatively bright sources located also outside the contamination area considered in the pipeline (Sect.~\ref{sec:contam}). Also nullified were the {\tt grvs\_mag} measurements of potential SB2 stars (i.e. for which double lines are detected in at least 10\% of epoch spectra), potential emission-line stars (i.e. with emission lines detected in more than 30\% of epoch spectra), and stars with S/N\,$<2$ in the mean spectrum over all transits. On the other hand, {\tt radial\_velocity} filters also removed stars with potentially good-quality {\tt grvs\_mag} measurements, such as hot and cool stars, for which radial velocity measurements were deemed insufficiently accurate. These stars are selected based on their effective temperature ($T_{\rm eff}$) stored in the parameter {\tt rv\_template\_teff}.\footnote{\label{fn:teff} {\tt rv\_template\_teff} is the name of the DR3-archive column containing the effective temperature of the synthetic spectrum associated to the star. The spectrum is selected from a synthetic spectral library (see the online documentation: \citealt{CU6-DR3-documentation}, section 6.2.3.3) based on the minimum distance to the assumed star atmospheric parameters.} Specifically,  {\tt grvs\_mag} was nullified for faint hot stars with ${\tt grvs\_mag}>12$~mag and $T_{\rm eff}\ge7000$~K;  for bright hot stars with ${\tt grvs\_mag}\le12$~mag and $T_{\rm eff}>14,500$~K; and for cool stars with $T_{\rm eff}< 3100$~K.

Additional filters were applied to identify and nullify other spurious estimates of {\tt grvs\_mag}. This includes about $9.4\times10^5$ stars with too few $G^{\rm{epoch}}_\mathrm{RVS}$ measurements, that is, with ${\tt grvs\_mag\_nbtransits}<3$ for faint stars with ${\tt grvs\_mag}\ge13$~mag, and ${\tt grvs\_mag\_nbtransits}~<2$ for brighter stars; about $1.5\times10^5$ stars fainter than ${\tt grvs\_mag}=14.1$~mag, corresponding to the faintest magnitude measurable in a single CCD spectrum given the uncertainties in background-flux estimation (Sect.~\ref{sec:background}); and another $\sim4.4\times10^4$ stars for which a flagging error in the pipeline procedure described in Sect.~\ref{sec:contam} prevented identification of a bright contaminating source. This error ---which led to a bright contaminant being ignored when a faint contaminant was also found--- affected only bright targets with ${\tt grvs\_mag}<12$~mag. To account for this, and given that stars with ${\tt grvs\_mag}\leq10.5$~mag rarely have bright contaminants, all stars fainter than 10.5~mag and with all their transits flagged `faint contaminant' had their {\tt grvs\_mag} nullified.

\begin{figure}[!ht]
  \begin{center}
  \includegraphics[width=8.5cm]{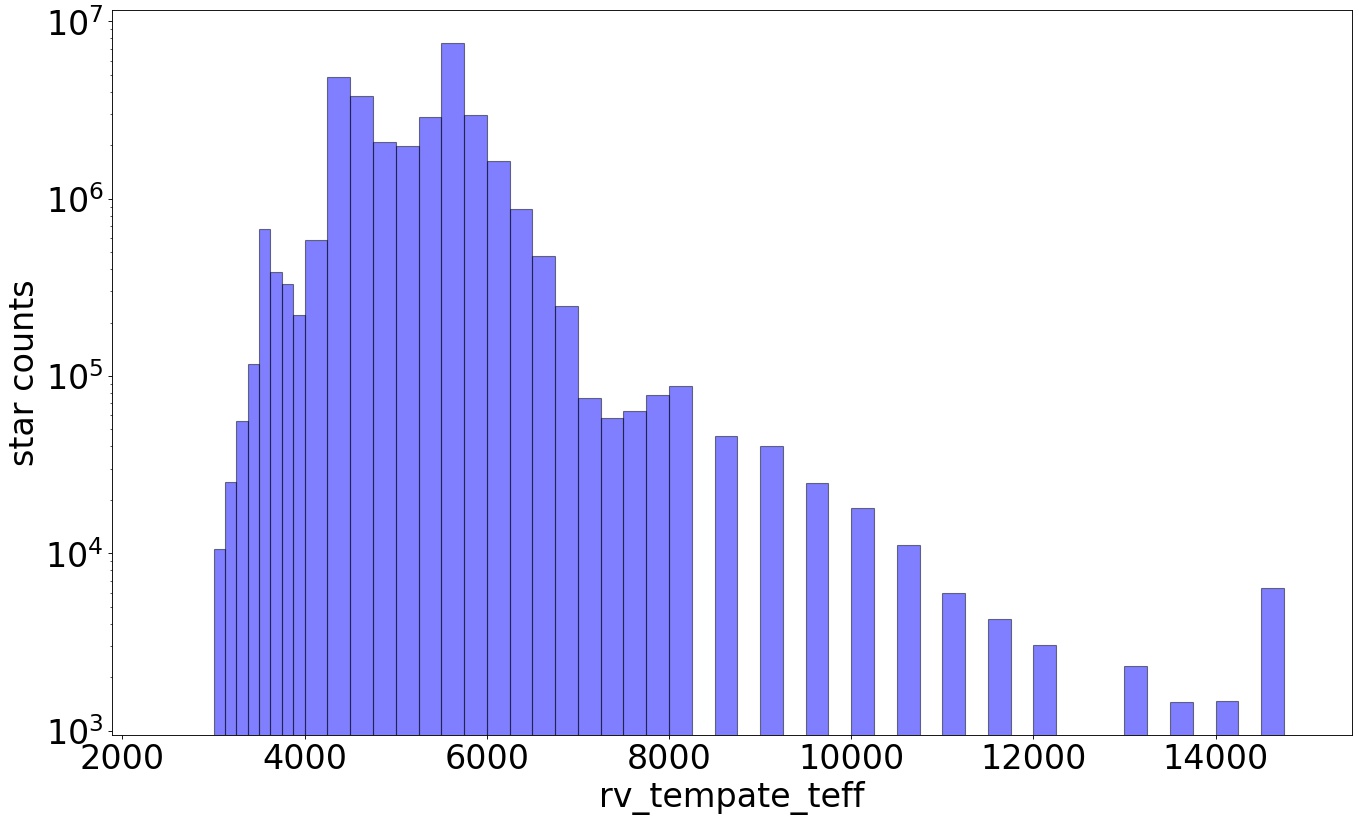}
 \end{center}
 \caption{Distribution of effective temperature ($T_{\rm eff}$, estimated as described in footnote~\ref{fn:teff}) for the 32.2 million stars with {\tt grvs\_mag} measurements published in DR3. The minimum $T_{\rm eff}$ is 3100~K, the maximum is 14~500~K, the median is 5250~K, and the mean is 5097~K.}
  \label{fig:teff}
 \end{figure}

\begin{figure}[!ht]
  \begin{center}
  \includegraphics[width=8.5cm]{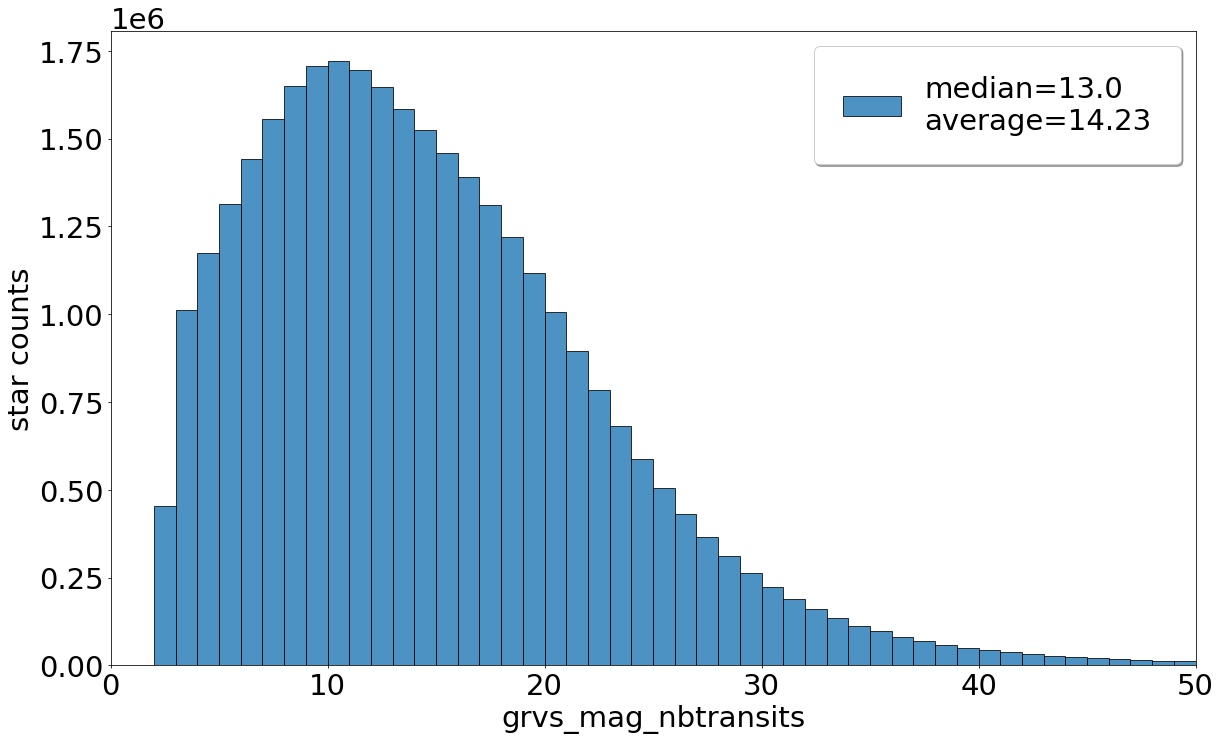}
 \end{center}
 \caption{Distribution of the number of epoch observations ({\tt grvs\_mag\_nbtransit}) for the 32.2 million stars with {\tt grvs\_mag} measurements published in DR3. The minimum {\tt grvs\_mag\_nbtransit} is 2, the maximum is 219, the median is 13, and the mean is 14.23. About 58~000 sources have ${\tt grvs\_mag\_nbtransit}>50$.}
 \label{fig:ntransits}
 \end{figure}

\begin{figure}[!ht]
 \begin{center}
\includegraphics[width=7.5cm]{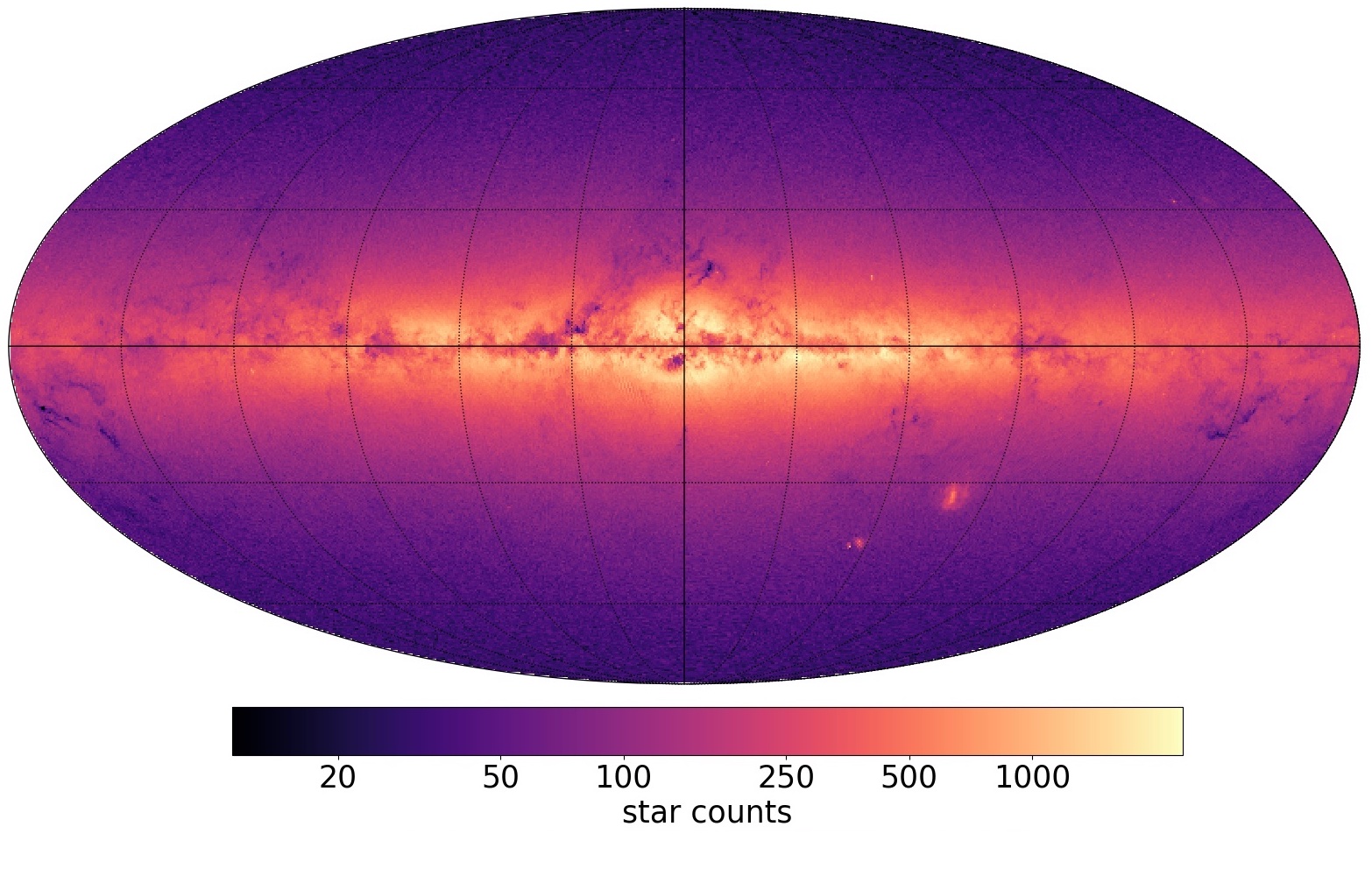}
\includegraphics[width=7.5cm]{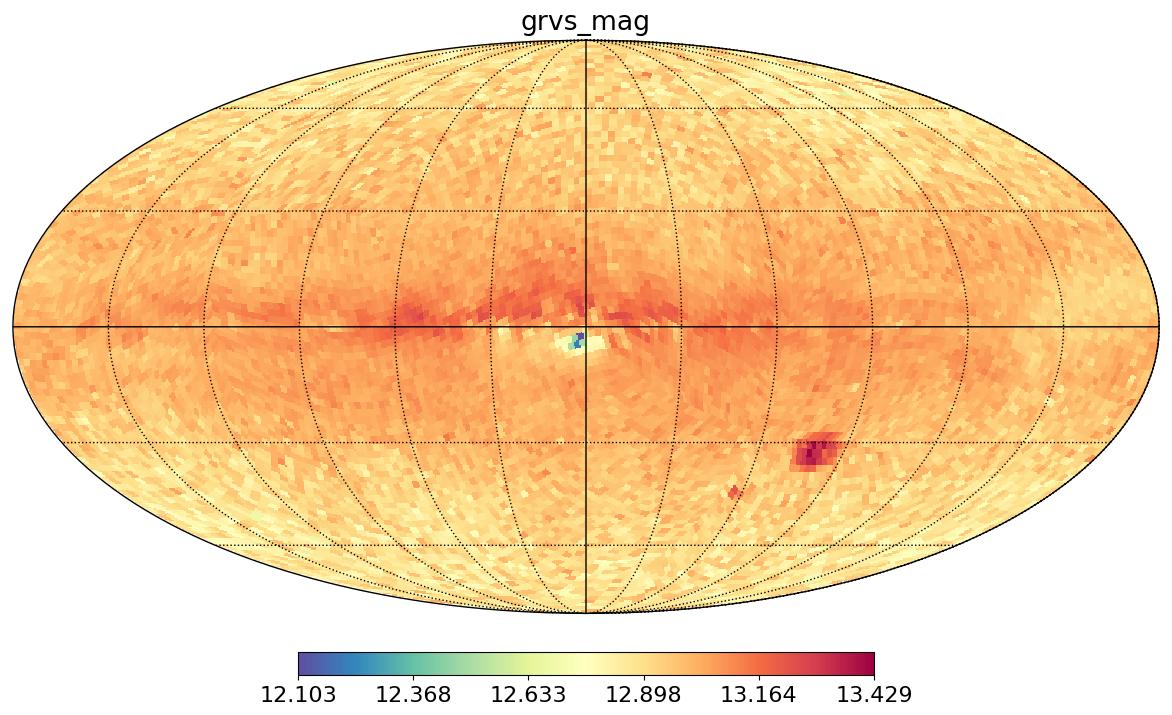}
\includegraphics[width=7.5cm]{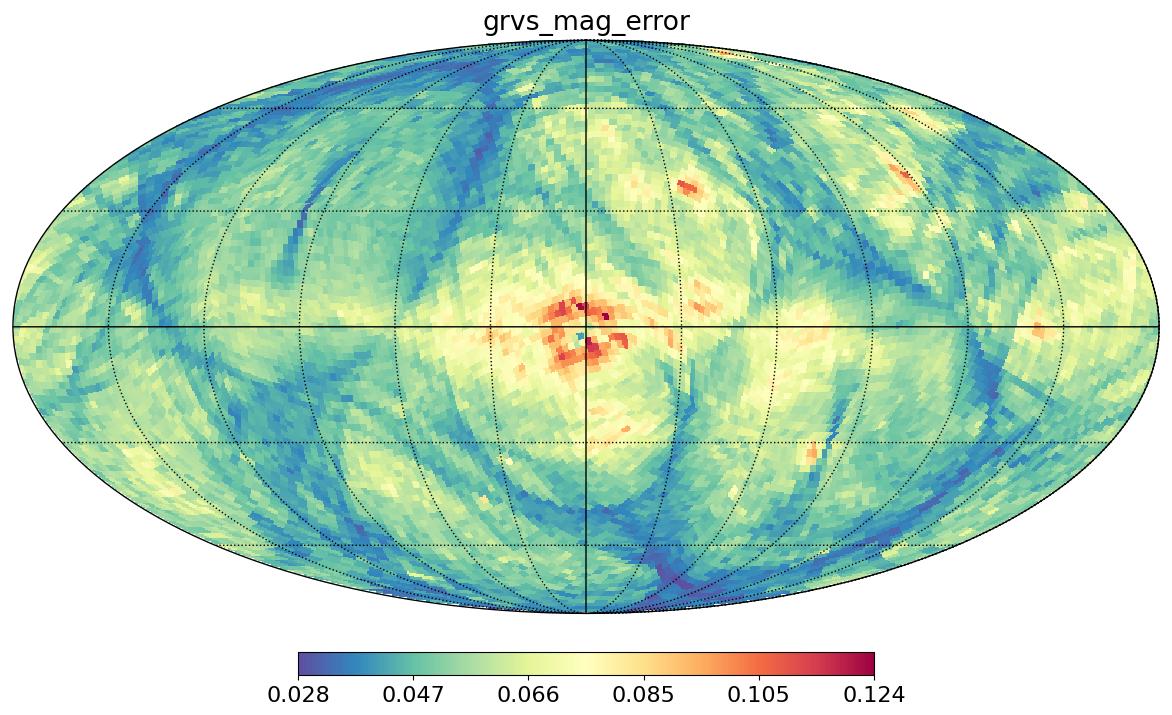}
\includegraphics[width=7.5cm]{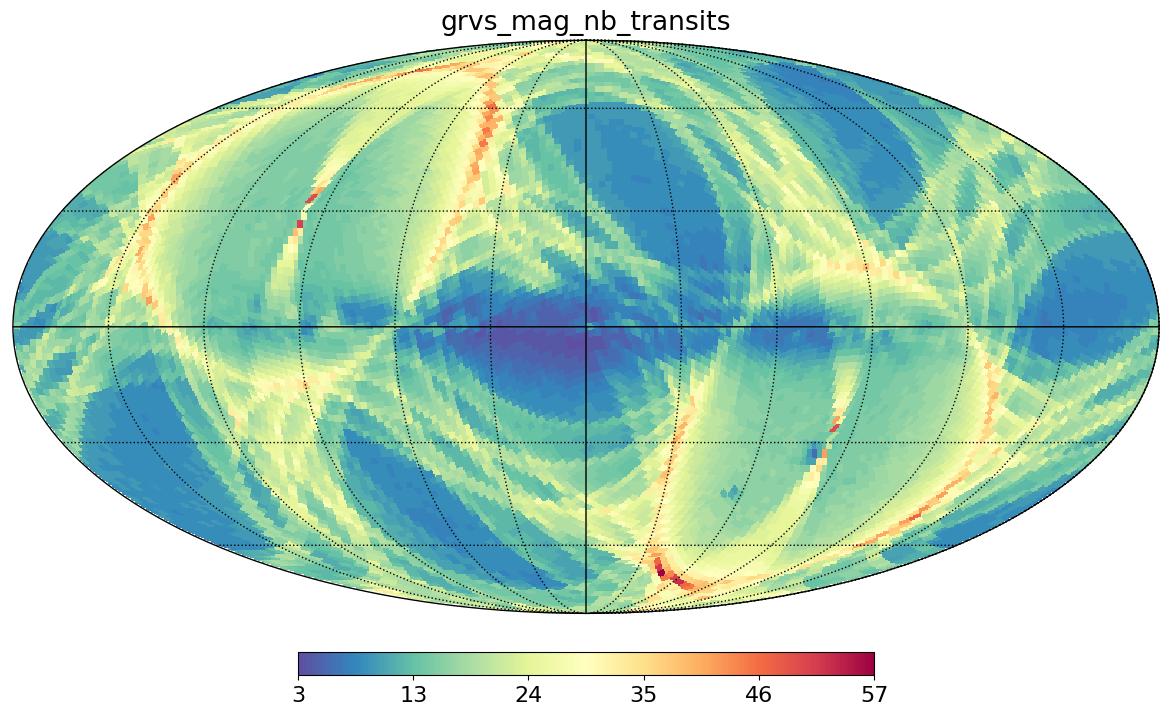}
 \end{center}
\caption{HEALPix maps in Galactic coordinates for the 32.2 million stars with {\tt grvs\_mag} measurements published in DR3. {\it From top to bottom:}  Source density over 0.2 square degrees (i.e. HEALPix level 7);  median {\tt grvs\_mag} over 3.36 square degrees (HEALPix level 5);  median {\tt grvs\_mag\_error} over 3.36 square degrees (HEALPix level 5); and  median {\tt grvs\_mag\_nb\_transits} over 3.36 square degrees (HEALPix level5). 
\label{fig:galmaps}}
\end{figure}

Overall, about 3.8 million {\tt grvs\_mag} measurements were nullified by the radial velocity filters, and another 1.1 million by the above additional {\tt grvs\_mag} filters. The vast majority of these 4.9 million cases (93\%) pertain to faint stars, with roughly 11.5\% of all stars fainter than ${\tt grvs\_mag}=12$ removed, compared to $\sim3.6$\% of brighter stars. For reference, we show in Figs.~\ref{fig:teff} and \ref{fig:ntransits} the distributions of effective temperature ($T_{\rm eff}$) and number of epoch measurements ({\tt grvs\_mag\_nbtransits}) for the 32.2 million stars with {\tt grvs\_mag} measurements.
Figure~\ref{fig:galmaps} (bottom panel) shows the sky distribution of {\tt grvs\_mag\_nbtransit}, the median number of epoch measurements per source being 13. 

While we expect most spurious {\tt grvs\_mag} measurements to have been cleaned through pipeline processing and the above validation procedure, some published measurements of faint stars with few observations, that is, with ${\tt grvs\_mag\_nbtransits}\simlt5$, could still be potentially spurious. A small {\tt grvs\_mag\_nbtransits} is typical for stars in crowded sky regions (Sect.~\ref{sec:complete}), and the few remaining observations may be affected by contamination that has gone unnoticed (the fainter the star, the fainter and more numerous the potential contaminants,  such as sufficiently bright stars outside the contamination area; see Sect.~\ref{sec:contam}). Faint stars are also significantly affected by background estimation errors, which are not averaged out when epoch observations are few. 

%%%%%%%%%%%%%%%%%%%%%%%%%%%%%%%%%%%%%%%%%%%%%%%%%%%%%%%%%%
%%%%%%%%%%%%%%%%%%%%%%%%%%%%%%%%%%%%%%%%%%%%%%%%
\subsection{Completeness of {\tt grvs\_mag}}\label{sec:complete}

The maps of Fig.~\ref{fig:galmaps} provide an illustration of the completeness of {\tt grvs\_mag} measurements across the sky. The sky distribution of the 32.2 million sources with DR3 {\tt grvs\_mag} measurements (top panel) shows that, unsurprisingly, the densest sampling is achieved in the Galactic plane and the two Magellanic clouds. The darker areas on this map correspond to regions obscured by dust lanes and to regions with extremely high stellar density, where clean {\tt grvs\_mag} measurements are particularly limited by strong contamination (Sect.\ref{sec:contam}), in addition to the constraint on the maximum number of spectra that can be obtained simultaneously (Sect.~\ref{sec:rvsspectra}).
The brightest areas on this map correspond to regions with high density of stars with {\tt grvs\_mag} measurements, but in such regions, the number of epoch measurements ({\tt grvs\_mag\_nbtransits}) is also lower, as revealed by the dark areas in the bottom map of Fig.~\ref{fig:galmaps} (the bright structures on this map show the imprints of the \gaia\ scanning law).
This is again because of the increased contamination and the limit to the number of RVS spectra that can be acquired in crowded areas. In such cases, priority is given to bright sources (Sect.~\ref{sec:rvsspectra}). For this reason, only the very few, brightest stars are observed close to the Galactic centre, and the median magnitude of this region, shown in the second map of Fig.~\ref{fig:galmaps}, is brighter than the average.

\begin{figure}[!ht]
  \begin{center}
  \includegraphics[width=8.5cm]{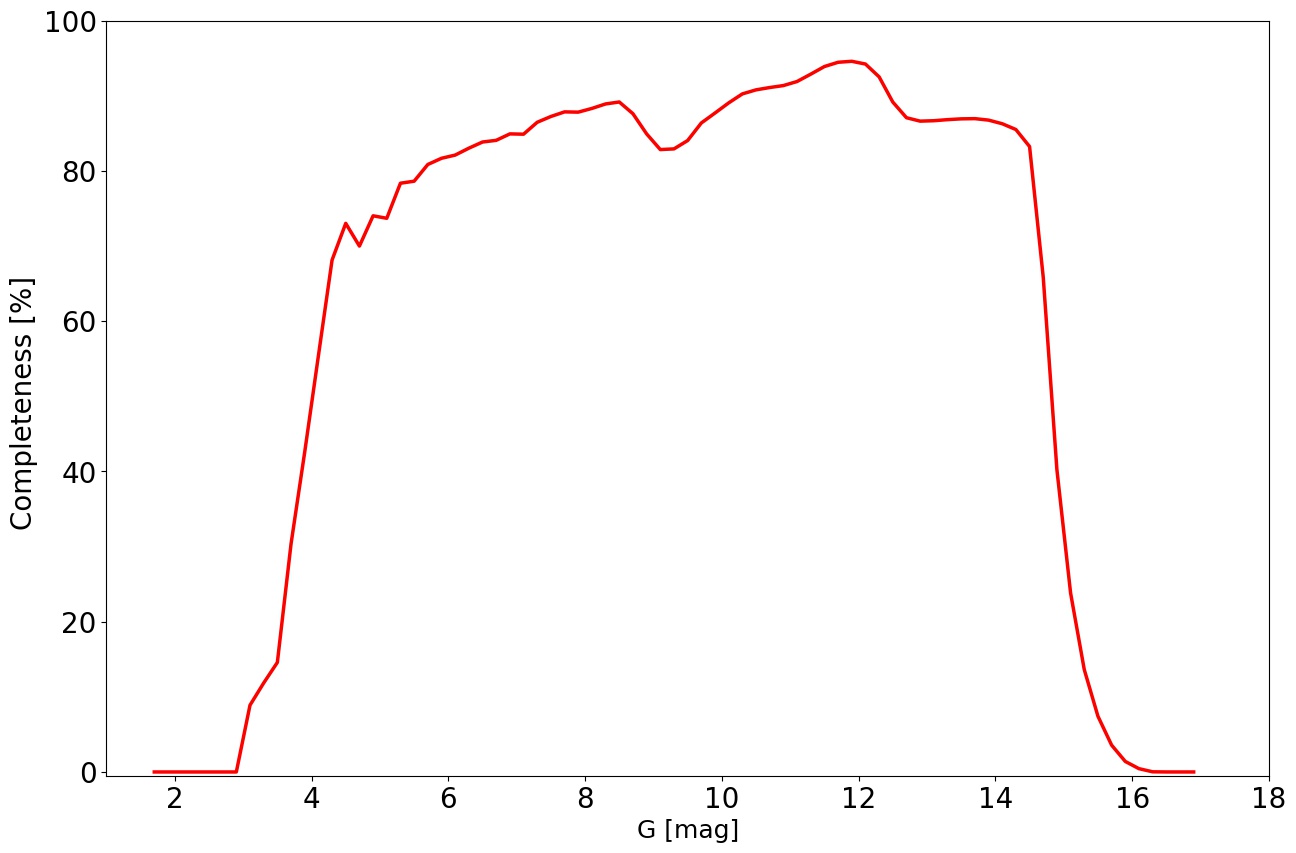}
 \end{center}
 \caption{Ratio of the number of stars with {\tt grvs\_mag} measurements to that of stars with standard \g\ measurements, in bins of 0.1 \g\ magnitude.}
  \label{fig:completeness}
 \end{figure}

We can estimate the completeness of {\tt grvs\_mag} measurements as the ratio of the number of stars with such measurements to that of stars with standard \g\ measurements. We compute this ratio in bins of 0.1 \g\ magnitude and present the results in Fig.~\ref{fig:completeness}. The completeness is better than 80\% over the full range of \g\ magnitudes from roughly 6 to 14~mag. Different features in the curve can be traced back to the procedure used to compute {\tt grvs\_mag}. The low completeness at the bright end ($\g<4$~mag) results from the removal of all saturated RVS spectra of bright stars (i.e. spectra presenting more than 40 saturated pixels). The relative drop in completeness at $\g\simgt8.5$~mag ($\obgrvs\simgt7$~mag) corresponds to the transition from 2D to 1D acquisition windows \citep[see figure~1 of][]{DR2-DPACP-47}. The 2D windows are never truncated or excluded (when not saturated), while 1D windows may be truncated and then excluded. A star around this transition may,  at some epoch, have \obgrvs slightly brighter than 7~mag and get a 2D window, and at some other epochs may have a slightly fainter magnitude and get a 1D window. In the latter case, a spike of the source PSF may occasionally cause the onboard software to assign an overlapping window to that spike (interpreted as a nearby source) and truncate the source window. The occurrence of such spurious detections drops sharply for faint stars \citep[see][section 3.2]{DR2-DPACP-47}, as shown by the rise in completeness at $\g \simgt 9$ in Fig.~\ref{fig:completeness}. The drop in completeness at $\g \simgt12.5$ (corresponding in median to ${\tt grvs\_mag}\sim12$) reflects the more effective filters applied to faint stars relative to bright stars (see Sect.~\ref{sec:filters}). 

We note that, because of the additional filters applied to {\tt grvs\_mag} measurements relative to {\tt radial\_velocity} measurements (Sect.~\ref{sec:epoch selection} and Sect.~\ref{sec:filters}), about 1.6 million sources with {\tt radial\_velocity} measurements do not have {\tt grvs\_mag} measurements.

%%%%%%%%%%%%%%%%%%%%%%%%%%%%%%%%%%%%%%%%%%%%%%%%%%%%%%%%%%%%%%%%
%%%%%%%%%%%%%%%%%%%%%%%%%%%%%%%%%%%%%%%%%%%%%%%%%%%%%%%%%%%%%%%%%

\subsection{Accuracy and precision}\label{sec:accuracy}

\begin{figure}[!ht]
  \begin{center}
  \includegraphics[width=8.5cm]{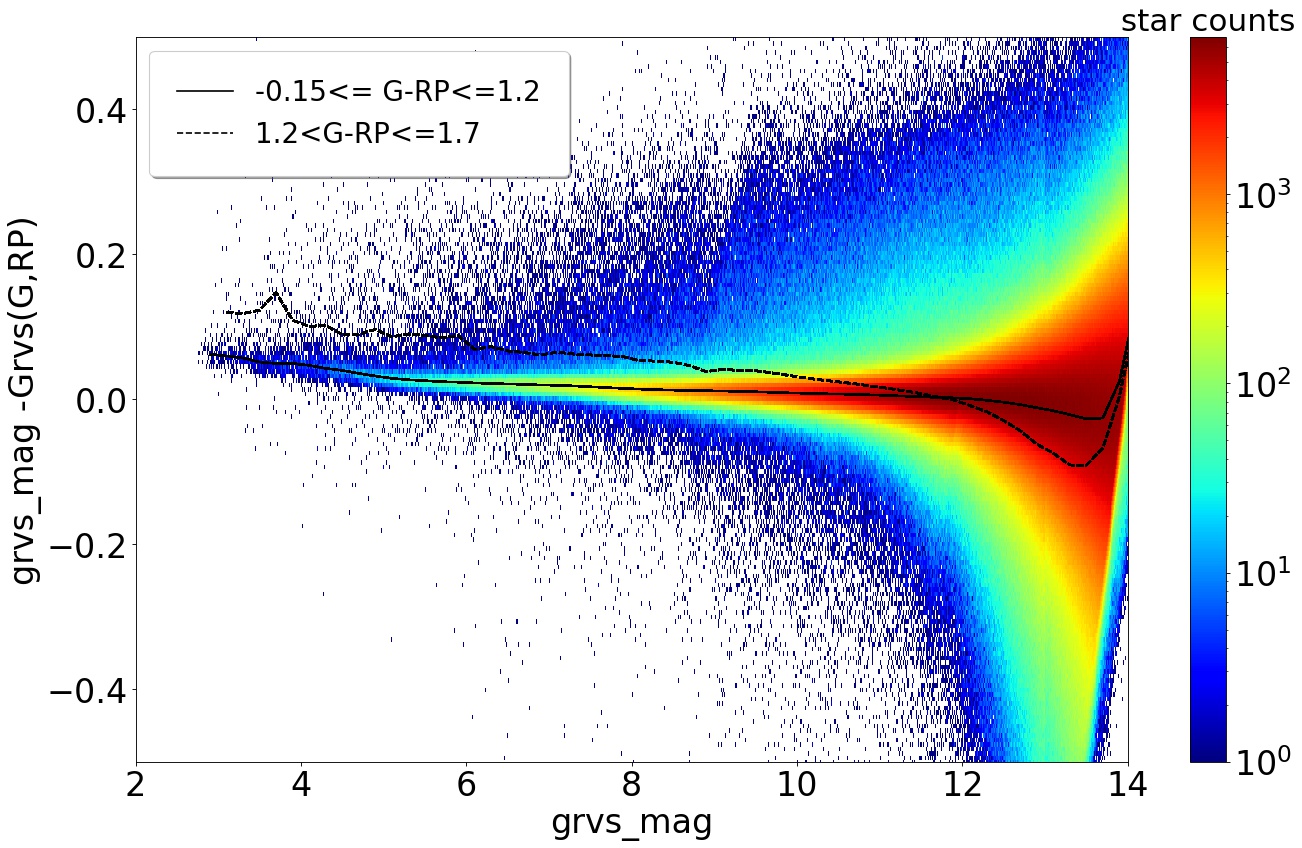}
   \end{center}
 \caption{${\tt grvs\_mag} - \grpgrvs$ residuals plotted against {\tt grvs\_mag}. The black solid line shows the median residuals in bins of $\Delta({\tt grvs\_mag})=0.1$~mag for stars with blue $G-G_{\rm RP}$ colours ($-0.15 \le G-G_{\rm RP} \le 1.2$~mag; Eq.~\ref{equ:bluerel}), and the black dashed line the median residuals for stars with red $G-G_{\rm RP}$ colours ($1.2<G-G_{\rm RP}\le 1.7$; Eq.~\ref{equ:redrel}). The $\extgrvs\le 14$~mag selection criterion to measure {\tt grvs\_mag} (Sect.~\ref{sec:selectextgrvs}) translates into a cut at ${\tt grvs\_mag} - \grpgrvs = {\tt grvs\_mag} - 14$ in this diagram, resulting in the apparent rise of the median residuals at the faint end. }
 \label{fig:accuracy}
 \end{figure}

We can estimate the accuracy (systematic uncertainty) of {\tt grvs\_mag} in first approximation by comparison with \grpgrvs\ (Sect.~\ref{sec:estimateCU5grvs}).
Figure~\ref{fig:accuracy} shows the ${\tt grvs\_mag} - \grpgrvs$ residuals as a function of {\tt grvs\_mag}. These residuals are sensitive to uncertainties in both {\tt grvs\_mag}  and \grpgrvs. 
The median residuals in bins of $\Delta({\tt grvs\_mag})=0.1$~mag in the blue range of $G-G_{\rm RP}$ colours (Eq.~\ref{equ:bluerel}; solid line in Fig.~\ref{fig:accuracy}) exhibit a small trend with {\tt grvs\_mag}, while the trend is more pronounced for the median residuals in the red range of colours (Eq.~\ref{equ:redrel}; dashed line). For both ranges of colours, the apparent sudden rise in median residuals at ${\tt grvs\_mag}\simgt13.75$~mag is artificial and caused by the selection criterion $\extgrvs\le 14$~mag (Sect.\ref{sec:selectextgrvs}). 

The large majority ($\sim$\,93\%) of the 32.2 million stars with {\tt grvs\_mag} measurements have blue $G-G_{\rm RP}$ colours, that is, $-0.15 \le G-G_{\rm RP} \le 1.2$~mag  (Sect.~\ref{sec:estimateCU5grvs}). The small trend in median ${\tt grvs\_mag} - \grpgrvs$ residuals in Fig.~\ref{fig:accuracy} is reminiscent of the trend observed 
in $\g-\grp$ versus \g\ \citep[see Fig.~32 of][]{2021A&A...649A...5F}.  A comparison with Hipparcos magnitude \citep{1997A&A...323L..49P} and Tycho2 colours \citep{2000A&A...355L..27H} reveals no saturation issue for {\tt grvs\_mag}. The saturation corrections for \g and \grp outlined in Appendix~C.1 of \citealt{2021A&A...649A...3R} do not reduce the trend. 
This is why the higher residuals at the bright end (where {\tt grvs\_mag} is slightly fainter than \grpgrvs) seem mostly imputable to systematic errors in \grpgrvs. Instead, the drop in residuals for faint stars with $12.2\simlt{\tt grvs\_mag}\simlt13.8$~mag (where {\tt grvs\_mag} is brighter than \grpgrvs) is mostly caused by systematic errors in {\tt grvs\_mag} introduced by the rejection of spectra with ${\it TotFlux}<0$ (see also Sect.~\ref{sec:grvssimu}). 

\begin{figure}[!ht]
  \begin{center}
  \includegraphics[width=8.5cm]{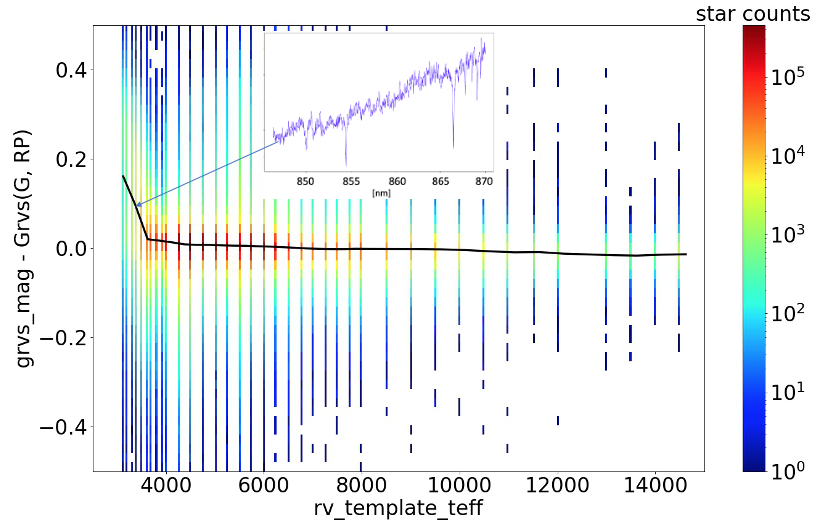}
  \end{center}
 \caption{${\tt grvs\_mag} - \grpgrvs$ residuals plotted against effective temperature $T_{\rm eff}$ ({\tt rv\_template\_teff}; see footnote~\ref{fn:teff}) for stars with ${\tt grvs\_mag}\le 12$~mag. The black line shows the median relation. For stars cooler than $T_{\rm eff}\sim3500$~K, {\tt grvs\_mag} is fainter than \grpgrvs. This is because of the positive gradient produced by TiO absorption in the spectra of such stars, as shown by the inset spectrum of a star with ${\tt grvs\_mag}=10.3$ and $T_{\rm eff}=3300$~K.}
 \label{fig:accuracyTeff}
 \end{figure}

The remaining 7\% ($\sim$\,2.2 million) of stars with {\tt grvs\_mag} measurements have red $G-G_{\rm RP}$ colours, that is, $1.2<G-G_{\rm RP}\le 1.7$ (Sect.~\ref{sec:estimateCU5grvs}). As seen in Fig.~\ref{fig:grvs_rp_vs_g_rp} above, the relation (Eq.~\ref{equ:redrel}) to estimate $G^{\rm{G,RP}}_\mathrm{RVS}$ from \g\ and $G_{\rm RP}$ in the red range of $G-G_{\rm RP}$ colours is not as well constrained as that (Eq.~\ref{equ:bluerel}) in the blue colour range, implying lower quality estimates of \grpgrvs in the red range.  
Indeed, stars with red $G-G_{\rm RP}$ colours tend to present few epoch observations (median ${\tt grvs\_mag\_nbtransits}=7$), characteristic of stars in dense regions (Sect.~\ref{sec:complete}). These stars are mostly distributed in the Galactic disk, where both stronger dust obscuration (the \g\ and $G_{\rm RP}$ filters extending over bluer wavelengths than the {\tt grvs\_mag} passband) and uncaught contamination (Sect.~\ref{sec:filters}) may contribute to making {\tt grvs\_mag} brighter than \grpgrvs. This is the reason for the pronounced trend in  median ${\tt grvs\_mag} - \grpgrvs$ residuals for faint stars around $12.2\simlt {\tt grvs\_mag}\simlt13.8$~mag in Fig.~\ref{fig:accuracy} (dashed line). Instead, at the bright end, the median residuals indicate that {\tt grvs\_mag} is systematically fainter than \grpgrvs, which reflects the prominence of cool stars ($T_{\rm eff}\simlt3500$~K) with positive gradients arising from the presence of TiO absorption in their spectra. This is further illustrated by Fig.~\ref{fig:accuracyTeff}, which shows the enhanced residuals affecting bright stars with $T_{\rm eff}\le3500$~K.

\begin{figure}[!ht]
  \begin{center}
  \includegraphics[width=8.5cm]{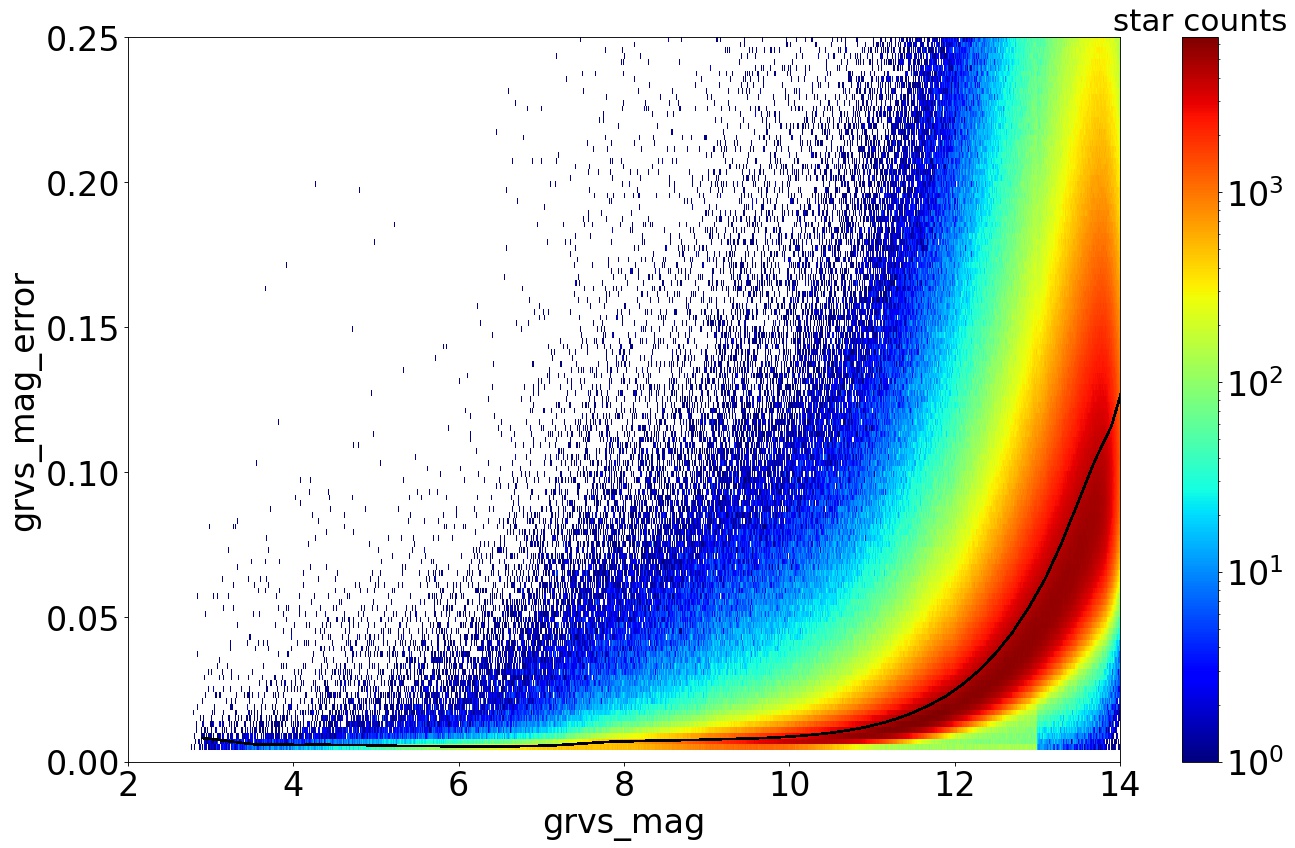}
  \end{center}
 \caption{Internal precision: {\tt grvs\_mag\_error} plotted against {\tt grvs\_mag}. The black line shows the median {\tt grvs\_mag\_error} in bins of $\Delta({\tt grvs\_mag})=0.1$~mag. This indicates a median ${\tt grvs\_mag\_error}$ around 0.006~mag over the range $4\simlt{\tt grvs\_mag}\simlt6.5$~mag, increasing to $\sim$\,0.025 at ${\tt grvs\_mag}\sim 12$~mag, and $\sim$\,0.125 at ${\tt grvs\_mag}\sim 14$~mag.}
 \label{fig:intprecision}
 \end{figure}

In Fig.~\ref{fig:intprecision}, we show the internal precision {\tt grvs\_mag\_error} as a function of {\tt grvs\_mag} for the 32.2 million stars with {\tt grvs\_mag} measurements. The distribution of {\tt grvs\_mag\_error} on the sky is shown in the third panel of Fig.~\ref{fig:galmaps}. A comparison with the distribution of the number of transits, {\tt grvs\_mag\_nbtransits} (bottom panel of Fig.~\ref{fig:galmaps}) indicates that, as expected, {\tt grvs\_mag\_error} tends to be larger when {\tt grvs\_mag\_nbtransits} is lower. 

%%%%%%%%%%%%%%%%%%%%%%%%%%%%%%%%%%%%%%%%%%%%%%%%%%%%%%%%%%%%%%%%%%%%%%
%                                                                    %
% Section 5 Estimate the mean GRVS filter transmission %
%                                                                    %
%%%%%%%%%%%%%%%%%%%%%%%%%%%%%%%%%%%%%%%%%%%%%%%%%%%%%%%%%%%%%%%%%%%%%%

\section{Estimation of the \grvs\ passband }\label{sec:grvsFilter}

We can estimate the passband of the effective \grvs\ filter to which {\tt grvs\_mag} corresponds by comparing high-quality RVS spectra acquired for stars with existing reference spectra. We consider 21 stars from the Hubble Space Telescope archive of calibrated spectrophotometric standards (CALSPEC) \citep[][2021 March Update]{2014PASP..126..711B} and 87 stars from the Next Generation Spectral Library (NGSL) spectrophotometric library \citep{2016ASPC..503..211H} which all have high-quality RVS data, that is, with ${\tt grvsmag\_error}<0.02$~mag and clean mean spectra in the 846--870~nm wavelength range (Fig.~\ref{fig:spectrum}). % = no NaN value 
To account for temporal variations of the zero-point (Sect.~\ref{sec:grvszp}), we scale each RVS epoch spectrum according to the zero-point estimated at the time of observation, corresponding to a factor $10^{-0.4~ZP(t_{obs})}$. We ignore the lower quality epoch spectra obtained before the first decontamination (i.e. 1317 OBMT, Fig.~\ref{fig:grvszptrend}).  
%The spectra are scaled back to electrons per second ($F_\mathrm{rvs}(\lambda)$) using the median zero point (21.3173). % filter_res
For each source, we then compute the spectrum averaged over all epochs, which we scale back to units of \es using the median zero-point ($ZP=21.317$). In the resulting spectrum, denoted $F_\mathrm{rvs}(\lambda)$, the total flux in the 846--870~nm wavelength range is therefore $10^{-0.4 ({\tt grvs\_mag}- ZP)}$. % filter_res_rescaled.

We seek the total RVS transmission $S(\lambda)$ such that
\begin{equation}
 F_\mathrm{rvs}(\lambda) = P F_\mathrm{ext}(\lambda) S(\lambda)\,,
\end{equation}
where $P$ is the telescope pupil area (0.7278~m$^2$) and $F_\mathrm{ext}(\lambda)$ the reference flux-calibrated spectrum of the source (from CALSPEC or NGSL; converted to units of photons~s$^{-1}$~m$^{-2}$).

To compare the RVS spectrum with the reference one, the RVS spectrum must be convolved to the (always lower) spectral resolution of the reference spectrum, and the reference spectrum must be  shifted to the RVS radial velocity reference frame. To do so, the optimal Gaussian kernel width and radial velocity shift are selected through a minimum $\chi^2$ search in the 848--870~nm wavelength range, corresponding to the quasi-flat range of the RVS transmission. The Gaussian kernel found in this way incorporates the blurring caused by temporal variations of the radial velocity. Then, the RVS transmission can be estimated by simply dividing the RVS spectrum by the reference spectrum. In practice, to avoid unwanted border effects, the RVS spectrum must be divided by a first-guess transmission (taken to be the nominal pre-launch one) before convolution to the reference-spectrum resolution, and then re-multiplied by the same transmission. The process is iterative, the spectra of all sources being processed at each iteration (convergence is obtained after three iterations). After the first iteration, the transmission over the full RVS wavelength range is estimated as the median transmission over all sources. The \grvs\ filter passband pertains to the 846--870~nm wavelength range used to compute \grvs. It is estimated through a B-spline median regression \citep[cobs R package,][]{cobs}, applying weights derived from the propagated errors of both the RVS and reference spectra. We note that five spectra found to be 5$\sigma$ outliers in their total flux were discarded. 

\begin{figure}[!ht]
  \begin{center}
  \includegraphics[width=8.5cm]{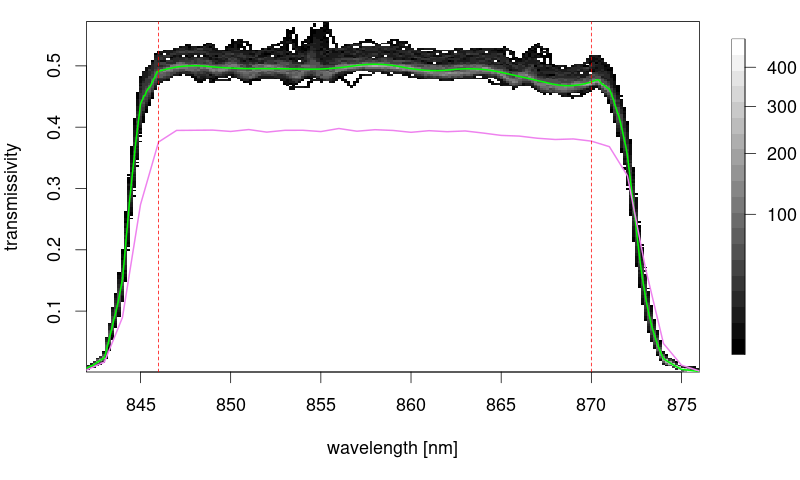}
 \end{center}
 \caption{Global RVS transmission function $S(\lambda)$ (green line) corresponding to the median of 108 estimates derived from reference spectra (shown in the background with a grey scale indicating curve density). The purple curve shows the nominal pre-launch transmission, and the two red vertical lines the limits of the \grvs\ passband. }
 \label{fig:fullfilter}
 \end{figure}

The full transmission $S(\lambda)$ is presented in Fig.~\ref{fig:fullfilter}, together with the nominal pre-launch transmission. This shows that the observed transmission is better than the pre-launch estimate (by a factor of about 1.23) and slightly shifted to the blue. The \grvs\ filter passband, which corresponds to this transmission in the 846--870~nm wavelength range, is available from the cosmos pages in footnote~\ref{fn:pb}.
%\footnote{\url{https://www.cosmos.esa.int/web/gaia/dr3-passbands}}

 By construction, the median zero-point to use with this filter is $ZP=21.317\pm0.002$~mag (see above).
This is consistent with the zero-point derived from the magnitude $m_0$ of the Vega spectrum $f_\lambda^{Vg}(\lambda)$ adopted for the \gbp and \grp zero-point determination,\footnote{\url{https://gea.esac.esa.int/archive/documentation/GEDR3/Data_processing/chap_cu5pho/cu5pho_sec_photProc/cu5pho_ssec_photCal.html}} converted to units of photons~s$^{-1}$~nm$^{-1}$~m$^{-2}$,

 \begin{equation}
 m_0 = 2.5 \log{\int P f_\lambda^{Vg}(\lambda) S(\lambda)  d\lambda}\,,
 \end{equation}
 which leads to $m_0 = 21.321\pm0.016$. % filter_res_rescaled

%%%%%%%%%%%%%%%%%%%%%%%%%%%%%%%%%%%%%%%%%%%%%%%%%%%%%%%%%%%%%%%%%%%%%%
%                                                                    %
% Section 8 Some possible usage of grvs_mag %
%                                                                    %
%%%%%%%%%%%%%%%%%%%%%%%%%%%%%%%%%%%%%%%%%%%%%%%%%%%%%%%%%%%%%%%%%%%%%%
\section{Examples of useful applications of {\tt grvs\_mag}}\label{sec:grvsUse}

By design, the present study, which is dedicated to the description of {\tt grvs\_mag} published in DR3, belongs to the series of `\gaia-processing papers', which reserve the scientific exploitation of DR3 data to the user community. In this section, we briefly illustrate some of the performances of {\tt grvs\_mag} measurements, which may be of interest for potential scientific applications.

\subsection{Extinction}
\begin{figure}[!ht]
  \begin{center}
  \includegraphics[width=8cm]{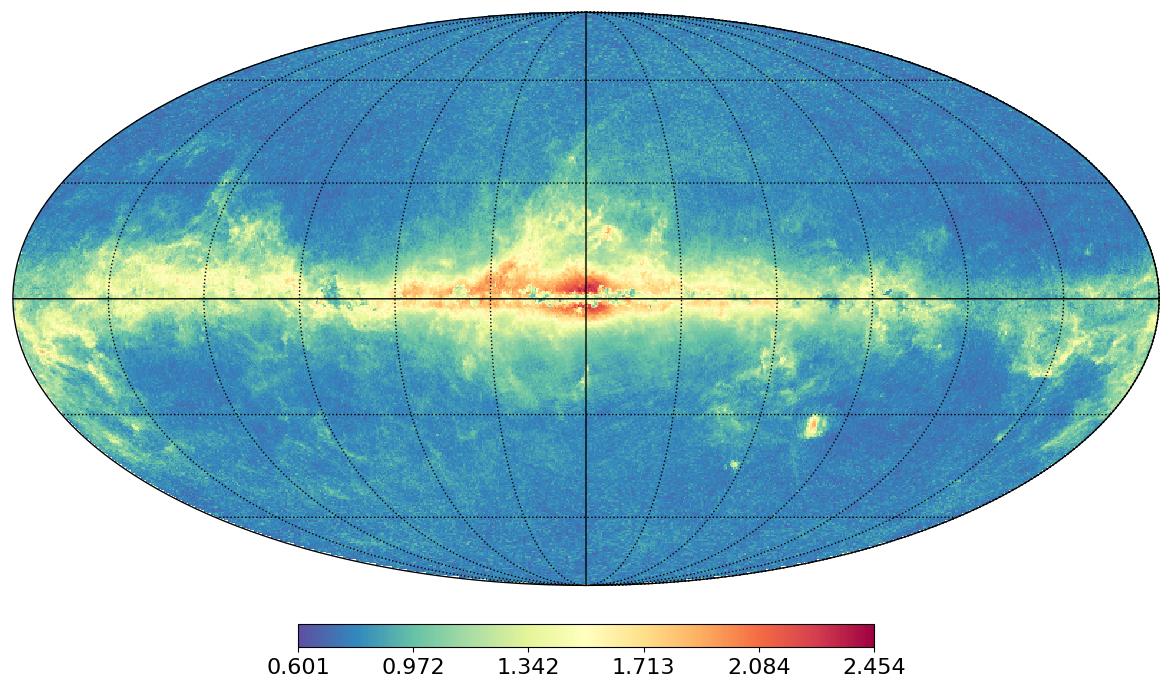}
 \end{center}
 \caption{Median $\g\ - {\tt grvs\_mag}$ colour in HEALPix map of level 7, highlighting the effect of extinction by interstellar dust.}
 \label{fig:galGminusGrvs}
 \end{figure}

The sky distribution of the $G - {\tt grvs\_mag}$ colours of the 32.2 million stars with DR3 {\tt grvs\_mag} measurements, shown in Fig.~\ref{fig:galGminusGrvs}, is sensitive to the distribution of extinction by interstellar dust.
The narrowness of the \grvs\ filter (Fig.~\ref{fig:passbands}) makes the extinction coefficient in this band mostly independent of the spectral type of the star and the extinction itself, unlike for the \gbp, \g,\ and \grp\ filters \citep[see e.g.][]{2010A&A...523A..48J}. Hence, {\tt grvs\_mag} is potentially useful for improving determinations of stellar atmospheric parameters. In fact, the derivation of these parameters from low-resolution, blue-, and red-photometer spectra suffers from a temperature--extinction degeneracy \citep{DR3-DPACP-156}, which  can be broken with help from an extra constraint in the red. 

\begin{figure}[!ht]
  \begin{center}
  \includegraphics[width=8cm]{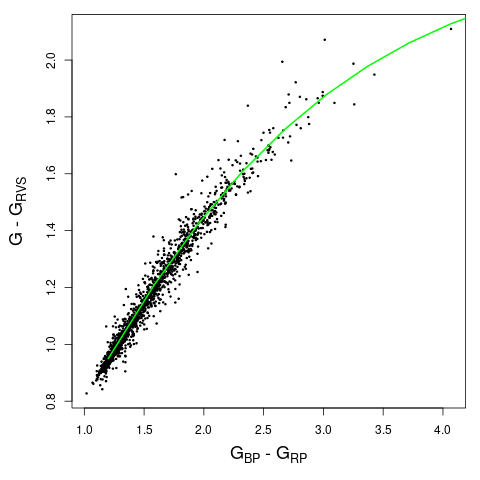}
 \end{center}
 \caption{$G - {\tt grvs\_mag}$ versus $G_{\rm BP}-G_{\rm RP}$ colour--colour diagram of APOGEE DR16 Red Clump stars with similar metallicity. The green line shows the expected red clump star colour--colour relation driven by extinction, as computed using the \cite{2019ApJ...886..108F} extinction law. }
 \label{fig:apogeeRCext}
 \end{figure}

Figure~\ref{fig:apogeeRCext} shows the $G - {\tt grvs\_mag}$ versus $G_{\rm BP}-G_{\rm RP}$ colour--colour relation of stars of red clump stars with similar metallicity ($\vert \mathrm{[M/H]} + 0.15 \vert< 2 \sigma_{\rm [M/H]}$) from the APOGEE DR16 Red Clump catalogue \citep{2014ApJ...790..127B}. This relation is primarily driven by extinction (we adopt here $G_\mathrm{BP}-G_\mathrm{RP}=1.2$ and $G-{\tt grvs\_mag}=0.95$ as the intrinsic Red Clump colours). Using the \cite{2019ApJ...886..108F} extinction law, we find that the extinction coefficient in the RVS band at the central wavelength of the \grvs\ filter is $k_\mathrm{RVS}=0.5385$. The extinction coefficients computed with the same extinction law for the $G$, \gbp, and \grp passbands are available from the cosmos pages.\footnote{\url{https://www.cosmos.esa.int/web/gaia/edr3-extinction-law}} As the passbands for these filters are much larger than the \grvs\ one (Fig.~\ref{fig:passbands}), their extinction coefficients depend sensitively on star colour and extinction itself, unlike $k_\mathrm{RVS}$ (which deviates by less than 0.2\% up to $A_0=20$). The green line in Fig.~\ref{fig:apogeeRCext} shows the resulting expected extinction effect on the $G - {\tt grvs\_mag}$ versus $G_{\rm BP}-G_{\rm RP}$ colours.

\subsection{Stellar metallicity}
\begin{figure}[!ht]
  \begin{center}
  \includegraphics[width=8.5cm]{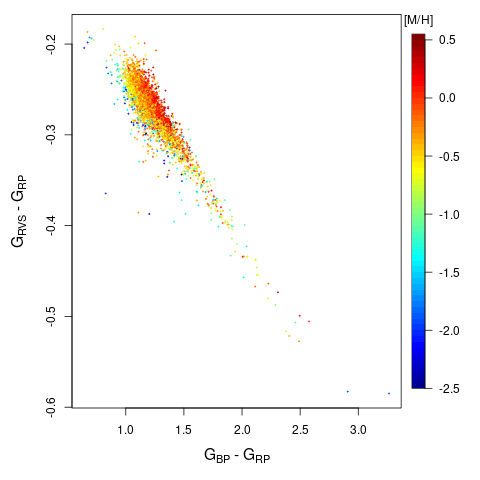}
   \end{center}
 \caption{${\tt grvs\_mag} - \grp$ versus  $\gbp\ - \grp$ colour--colour diagram for the APOGEE DR16 \citep{2020ApJS..249....3A} giants with low extinction ($A_0<0.05$, according to \citealt{2019A&A...625A.135L}). The points are colour-coded according to APOGEE metallicity [M/H].}
 \label{fig:giantsMH}
 \end{figure}

\begin{figure}[!ht]
  \begin{center}
  \includegraphics[width=7.5cm]{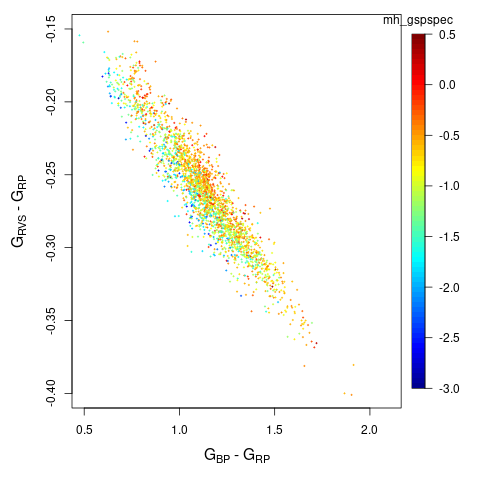}
   \end{center}
 \caption{${\tt grvs\_mag} - \grp$ versus $\gbp\ - \grp$ colour--colour diagram for the stars with $V_{\rm tot}>200$~\kms  in the Milky Way halo \citep{2018A&A...616A..10G}. The points are colour-coded according to metallicity {\tt mh\_gspspec} as published in DR3 \citep{DR3-DPACP-186}.}
 \label{fig:haloMH}
 \end{figure}

As noted in Fig.~\ref{fig:spectrum}, the \grvs narrow-band filter is centred on the infrared CaII triplet, which, in ground-based observations, is contaminated by atmospheric H$_2$O-line absorption. {\tt grvs\_mag} provides useful information complementary to that obtained from the \gaia\ blue- and red-prism spectra, which can help to constrain stellar atmospheric parameters. We note that \citet{2018A&A...611A..68B} predicted the potential of {\tt grvs\_mag} to constrain stellar metallicity. Figures~\ref{fig:giantsMH} and \ref{fig:haloMH} confirm the potential of the $\grp\ - {\tt grvs\_mag}$ colour as a metallicity diagnostic.
 
\subsection{Separating cool dwarfs from cool giants}
\begin{figure}[!ht]
  \begin{center}
  \includegraphics[width=7.5cm]{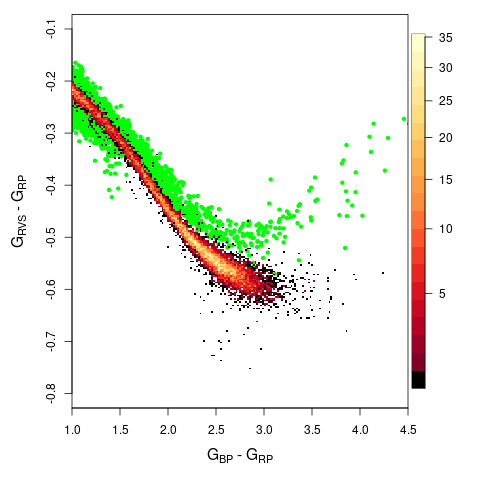}
   \end{center}
 \caption{${\tt grvs\_mag} - \grp$ versus $\gbp\ - \grp$ colour--colour diagram for cool stars in DR3. Low-extinction giants are represented with green dots, and nearby cool dwarfs with a red-coloured density plot.}
 \label{fig:coolstarsCC}
 \end{figure}

Figure~\ref{fig:coolstarsCC} shows how the ${\tt grvs\_mag}-G_{\rm RP}$ versus $G_{\rm BP}-G_{\rm RP}$ colour--colour relation differs between giants and red dwarfs among low-extinction cool stars, which can help disentangle the two populations. For this figure, all giants ($M_G=G+5+5 \log(\varpi/1000)>4$~mag) with low extinction ($A_0<0.05$, according to \citealt{2019A&A...625A.135L}) close to the Galactic plane ($|z|<500$~pc) have been used, while red dwarfs were selected simply with $\varpi>20$ and $M_G<4$~mag.
 
\section{Conclusions}\label{sec:conc}

We present the DR3 data, methodology, and validation procedure employed to compute the \grvs photometry published in \gaia\ DR3.
The \grvs photometry derived from RVS spectroscopy complements the \gaia\ photometry derived from astrometric and photometric data.
In particular, {\tt grvs\_mag} magnitudes, when combined with standard \g, \grp, and \gbp\ magnitudes, can improve astrophysical information on bright \gaia\ sources. We show examples of the potential of {\tt grvs\_mag} to constrain interstellar extinction and stellar metallicity and to separate cool dwarfs from cool giants.

The DR3 catalogue contains {\tt grvs\_mag} magnitudes ranging from 2.758 to 14.10~mag for 32.2 million stars with effective temperatures in the range $3100\simlt T_{\rm eff}\simlt14~500$~K. The median associated uncertainty, {\tt grvs\_mag\_error}, ranges from about 0.006 to 0.125~mag from the brightest to the faintest stars. Also listed in the catalogue is the number of epoch observations (or transits), {\tt grvs\_mag\_nbtransits}, used to compute {\tt grvs\_mag}. This ranges from 2 to 219, depending on the \gaia\ scanning law and  the density of the observed sky region (the densest regions allowing the fewest exploitable transits). The quantity {\tt grvs\_mag\_nbtransits} provides complementary information on the quality of {\tt grvs\_mag} measurements, since fewer transits in general lead to lower-quality data. Combined with {\tt grvs\_mag\_error}, it also provides an indication of the dispersion in  epoch magnitudes.

The {\tt grvs\_mag} magnitude recorded for each star in the DR3 catalogue is the median of all magnitudes obtained from the epoch observations of that star. The epoch magnitude is measured by integrating the flux in the cleaned RVS spectra, leaving out spectra deemed to be of poor quality, including spectra potentially contaminated by nearby sources and those with saturation issues.

The {\tt grvs\_mag\_error} uncertainty associated with {\tt grvs\_mag} is estimated as the formal error on the median, to which an error of 0.004~mag was added in quadrature to account for calibration-floor uncertainties. 

To estimate the passband of the effective \grvs\ filter to which {\tt grvs\_mag} corresponds (over the wavelength range from 846 to 870~nm), the spectra of 108 sources with both high-quality RVS spectra and reference spectra from the CALSPEC and NGSL spectrophotometric libraries were compared. The global RVS transmission derived in this way is better than the pre-launch estimate by a factor of about 1.23. The zero-point of this effective \grvs\ filter, calibrated based on the reference magnitudes of over $10^5$ constant Hipparcos stars, is $ZP=21.317\pm0.002$~mag. 

\gaia\ DR3 is an intermediate data release based on 34 months of mission data. The
next data release will be based on 66 months of data, with a correspondingly higher number of epoch observations and higher 
S/Ns for the combined data. 
 {\tt grvs\_mag} will be provided for stars fainter than 14.10~mag, and epoch magnitudes will also be published. 
Other novelties will include the improvement of systematic errors affecting faint stars, and the treatment of crowding.

\section*{Acknowledgements\label{sec:acknowl}}
\addcontentsline{toc}{chapter}{Acknowledgements}
This work presents results from the European Space Agency (ESA) space mission \gaia. \gaia\ data are being processed by the \gaia\ Data Processing and Analysis Consortium (DPAC). Funding for the DPAC is provided by national institutions, in particular the institutions participating in the \gaia\ MultiLateral Agreement (MLA). The \gaia\ mission website is \url{https://www.cosmos.esa.int/gaia}. The \gaia\ archive website is \url{https://archives.esac.esa.int/gaia}.
Acknowledgements are given in Appendix~\ref{ssec:appendixA}.


\bibliographystyle{aa} % style aa.bst
\bibliography{GAIA-CU6-GRVS} % refs.bib YOUR REFERENCES

\begin{appendix}
\section{}\label{ssec:appendixA}
This work presents results from the European Space Agency (ESA) space mission \gaia. \gaia\ data are being processed by the \gaia\ Data Processing and Analysis Consortium (DPAC). Funding for the DPAC is provided by national institutions, in particular the institutions participating in the \gaia\ MultiLateral Agreement (MLA). The \gaia\ mission website is \url{https://www.cosmos.esa.int/gaia}. The \gaia\ archive website is \url{https://archives.esac.esa.int/gaia}.

The \gaia\ mission and data processing have financially been supported by, in alphabetical order by country:
\begin{itemize}
\item the Algerian Centre de Recherche en Astronomie, Astrophysique et G\'{e}ophysique of Bouzareah Observatory;
\item the Austrian Fonds zur F\"{o}rderung der wissenschaftlichen Forschung (FWF) Hertha Firnberg Programme through grants T359, P20046, and P23737;
\item the BELgian federal Science Policy Office (BELSPO) through various PROgramme de D\'{e}veloppement d'Exp\'{e}riences scientifiques (PRODEX) grants and the Polish Academy of Sciences - Fonds Wetenschappelijk Onderzoek through grant VS.091.16N, and the Fonds de la Recherche Scientifique (FNRS), and the Research Council of Katholieke Universiteit (KU) Leuven through grant C16/18/005 (Pushing AsteRoseismology to the next level with TESS, GaiA, and the Sloan DIgital Sky SurvEy -- PARADISE);  
\item the Brazil-France exchange programmes Funda\c{c}\~{a}o de Amparo \`{a} Pesquisa do Estado de S\~{a}o Paulo (FAPESP) and Coordena\c{c}\~{a}o de Aperfeicoamento de Pessoal de N\'{\i}vel Superior (CAPES) - Comit\'{e} Fran\c{c}ais d'Evaluation de la Coop\'{e}ration Universitaire et Scientifique avec le Br\'{e}sil (COFECUB);
\item the Chilean Agencia Nacional de Investigaci\'{o}n y Desarrollo (ANID) through Fondo Nacional de Desarrollo Cient\'{\i}fico y Tecnol\'{o}gico (FONDECYT) Regular Project 1210992 (L.~Chemin);
\item the National Natural Science Foundation of China (NSFC) through grants 11573054, 11703065, and 12173069, the China Scholarship Council through grant 201806040200, and the Natural Science Foundation of Shanghai through grant 21ZR1474100;  
\item the Tenure Track Pilot Programme of the Croatian Science Foundation and the \'{E}cole Polytechnique F\'{e}d\'{e}rale de Lausanne and the project TTP-2018-07-1171 `Mining the Variable Sky', with the funds of the Croatian-Swiss Research Programme;
\item the Czech-Republic Ministry of Education, Youth, and Sports through grant LG 15010 and INTER-EXCELLENCE grant LTAUSA18093, and the Czech Space Office through ESA PECS contract 98058;
\item the Danish Ministry of Science;
\item the Estonian Ministry of Education and Research through grant IUT40-1;
\item the European Commission’s Sixth Framework Programme through the European Leadership in Space Astrometry (\href{https://www.cosmos.esa.int/web/gaia/elsa-rtn-programme}{ELSA}) Marie Curie Research Training Network (MRTN-CT-2006-033481), through Marie Curie project PIOF-GA-2009-255267 (Space AsteroSeismology \& RR Lyrae stars, SAS-RRL), and through a Marie Curie Transfer-of-Knowledge (ToK) fellowship (MTKD-CT-2004-014188); the European Commission's Seventh Framework Programme through grant FP7-606740 (FP7-SPACE-2013-1) for the \gaia\ European Network for Improved data User Services (\href{https://gaia.ub.edu/twiki/do/view/GENIUS/}{GENIUS}) and through grant 264895 for the \gaia\ Research for European Astronomy Training (\href{https://www.cosmos.esa.int/web/gaia/great-programme}{GREAT-ITN}) network;
\item the European Cooperation in Science and Technology (COST) through COST Action CA18104 `Revealing the Milky Way with \gaia (MW-Gaia)';
\item the European Research Council (ERC) through grants 320360, 647208, and 834148 and through the European Union’s Horizon 2020 research and innovation and excellent science programmes through Marie Sk{\l}odowska-Curie grant 745617 (Our Galaxy at full HD -- Gal-HD) and 895174 (The build-up and fate of self-gravitating systems in the Universe) as well as grants 687378 (Small Bodies: Near and Far), 682115 (Using the Magellanic Clouds to Understand the Interaction of Galaxies), 695099 (A sub-percent distance scale from binaries and Cepheids -- CepBin), 716155 (Structured ACCREtion Disks -- SACCRED), 951549 (Sub-percent calibration of the extragalactic distance scale in the era of big surveys -- UniverScale), and 101004214 (Innovative Scientific Data Exploration and Exploitation Applications for Space Sciences -- EXPLORE);
\item the European Science Foundation (ESF), in the framework of the \gaia\ Research for European Astronomy Training Research Network Programme (\href{https://www.cosmos.esa.int/web/gaia/great-programme}{GREAT-ESF});
\item the European Space Agency (ESA) in the framework of the \gaia\ project, through the Plan for European Cooperating States (PECS) programme through contracts C98090 and 4000106398/12/NL/KML for Hungary, through contract 4000115263/15/NL/IB for Germany, and through PROgramme de D\'{e}veloppement d'Exp\'{e}riences scientifiques (PRODEX) grant 4000127986 for Slovenia;  
\item the Academy of Finland through grants 299543, 307157, 325805, 328654, 336546, and 345115 and the Magnus Ehrnrooth Foundation;
\item the French Centre National d’\'{E}tudes Spatiales (CNES), the Agence Nationale de la Recherche (ANR) through grant ANR-10-IDEX-0001-02 for the `Investissements d'avenir' programme, through grant ANR-15-CE31-0007 for project `Modelling the Milky Way in the \gaia era’ (MOD4Gaia), through grant ANR-14-CE33-0014-01 for project `The Milky Way disc formation in the \gaia era’ (ARCHEOGAL), through grant ANR-15-CE31-0012-01 for project `Unlocking the potential of Cepheids as primary distance calibrators’ (UnlockCepheids), through grant ANR-19-CE31-0017 for project `Secular evolution of galxies' (SEGAL), and through grant ANR-18-CE31-0006 for project `Galactic Dark Matter' (GaDaMa), the Centre National de la Recherche Scientifique (CNRS) and its SNO \gaia of the Institut des Sciences de l’Univers (INSU), its Programmes Nationaux: Cosmologie et Galaxies (PNCG), Gravitation R\'{e}f\'{e}rences Astronomie M\'{e}trologie (PNGRAM), Plan\'{e}tologie (PNP), Physique et Chimie du Milieu Interstellaire (PCMI), and Physique Stellaire (PNPS), the `Action F\'{e}d\'{e}ratrice \gaia' of the Observatoire de Paris, the R\'{e}gion de Franche-Comt\'{e}, the Institut National Polytechnique (INP) and the Institut National de Physique nucl\'{e}aire et de Physique des Particules (IN2P3) co-funded by CNES;
\item the German Aerospace Agency (Deutsches Zentrum f\"{u}r Luft- und Raumfahrt e.V., DLR) through grants 50QG0501, 50QG0601, 50QG0602, 50QG0701, 50QG0901, 50QG1001, 50QG1101, 50\-QG1401, 50QG1402, 50QG1403, 50QG1404, 50QG1904, 50QG2101, 50QG2102, and 50QG2202, and the Centre for Information Services and High Performance Computing (ZIH) at the Technische Universit\"{a}t Dresden for generous allocations of computer time;
\item the Hungarian Academy of Sciences through the Lend\"{u}let Programme grants LP2014-17 and LP2018-7 and the Hungarian National Research, Development, and Innovation Office (NKFIH) through grant KKP-137523 (`SeismoLab');
\item the Science Foundation Ireland (SFI) through a Royal Society - SFI University Research Fellowship (M.~Fraser);
\item the Israel Ministry of Science and Technology through grant 3-18143 and the Tel Aviv University Center for Artificial Intelligence and Data Science (TAD) through a grant;
\item the Agenzia Spaziale Italiana (ASI) through contracts I/037/08/0, I/058/10/0, 2014-025-R.0, 2014-025-R.1.2015, and 2018-24-HH.0 to the Italian Istituto Nazionale di Astrofisica (INAF), contract 2014-049-R.0/1/2 to INAF for the Space Science Data Centre (SSDC, formerly known as the ASI Science Data Center, ASDC), contracts I/008/10/0, 2013/030/I.0, 2013-030-I.0.1-2015, and 2016-17-I.0 to the Aerospace Logistics Technology Engineering Company (ALTEC S.p.A.), INAF, and the Italian Ministry of Education, University, and Research (Ministero dell'Istruzione, dell'Universit\`{a} e della Ricerca) through the Premiale project `MIning The Cosmos Big Data and Innovative Italian Technology for Frontier Astrophysics and Cosmology' (MITiC);
\item the Netherlands Organisation for Scientific Research (NWO) through grant NWO-M-614.061.414, through a VICI grant (A.~Helmi), and through a Spinoza prize (A.~Helmi), and the Netherlands Research School for Astronomy (NOVA);
\item the Polish National Science Centre through HARMONIA grant 2018/30/M/ST9/00311 and DAINA grant 2017/27/L/ST9/03221 and the Ministry of Science and Higher Education (MNiSW) through grant DIR/WK/2018/12;
\item the Portuguese Funda\c{c}\~{a}o para a Ci\^{e}ncia e a Tecnologia (FCT) through national funds, grants SFRH/\-BD/128840/2017 and PTDC/FIS-AST/30389/2017, and work contract DL 57/2016/CP1364/CT0006, the Fundo Europeu de Desenvolvimento Regional (FEDER) through grant POCI-01-0145-FEDER-030389 and its Programa Operacional Competitividade e Internacionaliza\c{c}\~{a}o (COMPETE2020) through grants UIDB/04434/2020 and UIDP/04434/2020, and the Strategic Programme UIDB/\-00099/2020 for the Centro de Astrof\'{\i}sica e Gravita\c{c}\~{a}o (CENTRA);  
\item the Slovenian Research Agency through grant P1-0188;
\item the Spanish Ministry of Economy (MINECO/FEDER, UE), the Spanish Ministry of Science and Innovation (MICIN), the Spanish Ministry of Education, Culture, and Sports, and the Spanish Government through grants BES-2016-078499, BES-2017-083126, BES-C-2017-0085, ESP2016-80079-C2-1-R, ESP2016-80079-C2-2-R, FPU16/03827, PDC2021-121059-C22, RTI2018-095076-B-C22, and TIN2015-65316-P (`Computaci\'{o}n de Altas Prestaciones VII'), the Juan de la Cierva Incorporaci\'{o}n Programme (FJCI-2015-2671 and IJC2019-04862-I for F.~Anders), the Severo Ochoa Centre of Excellence Programme (SEV2015-0493), and MICIN/AEI/10.13039/501100011033 (and the European Union through European Regional Development Fund `A way of making Europe') through grant RTI2018-095076-B-C21, the Institute of Cosmos Sciences University of Barcelona (ICCUB, Unidad de Excelencia `Mar\'{\i}a de Maeztu’) through grant CEX2019-000918-M, the University of Barcelona's official doctoral programme for the development of an R+D+i project through an Ajuts de Personal Investigador en Formaci\'{o} (APIF) grant, the Spanish Virtual Observatory through project AyA2017-84089, the Galician Regional Government, Xunta de Galicia, through grants ED431B-2021/36, ED481A-2019/155, and ED481A-2021/296, the Centro de Investigaci\'{o}n en Tecnolog\'{\i}as de la Informaci\'{o}n y las Comunicaciones (CITIC), funded by the Xunta de Galicia and the European Union (European Regional Development Fund -- Galicia 2014-2020 Programme), through grant ED431G-2019/01, the Red Espa\~{n}ola de Supercomputaci\'{o}n (RES) computer resources at MareNostrum, the Barcelona Supercomputing Centre - Centro Nacional de Supercomputaci\'{o}n (BSC-CNS) through activities AECT-2017-2-0002, AECT-2017-3-0006, AECT-2018-1-0017, AECT-2018-2-0013, AECT-2018-3-0011, AECT-2019-1-0010, AECT-2019-2-0014, AECT-2019-3-0003, AECT-2020-1-0004, and DATA-2020-1-0010, the Departament d'Innovaci\'{o}, Universitats i Empresa de la Generalitat de Catalunya through grant 2014-SGR-1051 for project `Models de Programaci\'{o} i Entorns d'Execuci\'{o} Parallels' (MPEXPAR), and Ramon y Cajal Fellowship RYC2018-025968-I funded by MICIN/AEI/10.13039/501100011033 and the European Science Foundation (`Investing in your future');
\item the Swedish National Space Agency (SNSA/Rymdstyrelsen);
\item the Swiss State Secretariat for Education, Research, and Innovation through the Swiss Activit\'{e}s Nationales Compl\'{e}mentaires and the Swiss National Science Foundation through an Eccellenza Professorial Fellowship (award PCEFP2\_194638 for R.~Anderson);
\item the United Kingdom Particle Physics and Astronomy Research Council (PPARC), the United Kingdom Science and Technology Facilities Council (STFC), and the United Kingdom Space Agency (UKSA) through the following grants to the University of Bristol, the University of Cambridge, the University of Edinburgh, the University of Leicester, the Mullard Space Sciences Laboratory of University College London, and the United Kingdom Rutherford Appleton Laboratory (RAL): PP/D006511/1, PP/D006546/1, PP/D006570/1, ST/I000852/1, ST/J005045/1, ST/K00056X/1, ST/\-K000209/1, ST/K000756/1, ST/L006561/1, ST/N000595/1, ST/N000641/1, ST/N000978/1, ST/\-N001117/1, ST/S000089/1, ST/S000976/1, ST/S000984/1, ST/S001123/1, ST/S001948/1, ST/\-S001980/1, ST/S002103/1, ST/V000969/1, ST/W002469/1, ST/W002493/1, ST/W002671/1, ST/W002809/1, and EP/V520342/1.
\end{itemize}

The GBOT programme  uses observations collected at (i) the European Organisation for Astronomical Research in the Southern Hemisphere (ESO) with the VLT Survey Telescope (VST), under ESO programmes
092.B-0165,
093.B-0236,
094.B-0181,
095.B-0046,
096.B-0162,
097.B-0304,
098.B-0030,
099.B-0034,
0100.B-0131,
0101.B-0156,
0102.B-0174, and
0103.B-0165;
and (ii) the Liverpool Telescope, which is operated on the island of La Palma by Liverpool John Moores University in the Spanish Observatorio del Roque de los Muchachos of the Instituto de Astrof\'{\i}sica de Canarias with financial support from the United Kingdom Science and Technology Facilities Council, and (iii) telescopes of the Las Cumbres Observatory Global Telescope Network.
AMB acknowledges funding from the European Union's Horizon 2020 research and innovation programme under the Marie Sk\l{}odowska-Curie grant agreement No 895174. We made use of TOPCAT: \url{http://www.starlink.ac.uk/topcat}. 

\end{appendix}
\end{document}